\documentclass[a4paper,11pt]{article}
\pdfoutput=1 

\usepackage{jinstpub} 

\usepackage{color}
\definecolor{darkgreen}{RGB}{0,150,0}
\usepackage{hyperref}
\hypersetup{
  colorlinks   = true, 
 urlcolor     = darkgreen, 
  linkcolor    = blue, 
  citecolor   = blue 
}
\usepackage{psvectorian} 
\usepackage{cleveref} 
\usepackage{xfrac} 
\usepackage{threeparttable} 
\usepackage{placeins} 
\usepackage{amssymb} 
\usepackage[T1]{fontenc}
\usepackage[percent]{overpic}
\usepackage[left,modulo]{lineno}
\usepackage{upgreek} 

\crefname{figure}{figure}{figures}
\crefname{Figure}{Figure}{Figures}
\crefname{equation}{eq.}{eqs.}
\crefname{Equation}{Eq.}{Eqs.}

\begin{document}

\title{Position Reconstruction in LUX}

\author{LUX collaboration.}
\author[a,b,c]{D.S.~Akerib,}
\author[d]{S.~Alsum,}  
\author[e]{H.M.~Ara\'{u}jo,} 
\author[f]{X.~Bai,}  
\author[e]{A.J.~Bailey,} 
\author[g]{J.~Balajthy,}  
\author[h]{P.~Beltrame,} 
\author[i,j]{E.P.~Bernard,} 
\author[k]{A.~Bernstein,}   
\author[a,b,c]{T.P.~Biesiadzinski,} 
\author[i,j]{E.M.~Boulton,}
\author[l]{P.~Br\'as,}  
\author[m,n]{D.~Byram,} 
\author[j]{S.B.~Cahn,} 
\author[o,p]{M.C.~Carmona-Benitez,} 
\author[q]{C.~Chan,}  
\author[e]{A.~Currie,} 
\author[r]{J.E.~Cutter,}   
\author[h]{T.J.R.~Davison,} 
\author[s]{A.~Dobi,} 
\author[t]{E.~Druszkiewicz,} 
\author[j]{B.N.~Edwards,} 
\author[u]{S.R.~Fallon,}  
\author[b,c]{A.~Fan,}
\author[q,s]{S.~Fiorucci,} 
\author[q]{R.J.~Gaitskell,}  
\author[u]{J.~Genovesi,} 
\author[v]{C.~Ghag,}  
\author[s]{M.G.D.~Gilchriese,}
\author[g]{C.R.~Hall,}   
\author[f,n]{M.~Hanhardt,} 
\author[p]{S.J.~Haselschwardt,}
\author[j,s,w]{S.A.~Hertel,} 
\author[i]{D.P.~Hogan,}
\author[i,j,n]{M.~Horn,} 
\author[q]{D.Q.~Huang,} 
\author[b,c]{C.M.~Ignarra,} 
\author[i]{R.G.~Jacobsen,} 
\author[a,b,c]{W.~Ji,} 
\author[i]{K.~Kamdin,}
\author[k]{K.~Kazkaz,} 
\author[t]{D.~Khaitan,} 
\author[g]{R.~Knoche,} 
\author[j]{N.A.~Larsen,} 
\author[k,r]{B.G.~Lenardo,} 
\author[s]{K.T.~Lesko,} 
\author[l]{A.~Lindote,} 
\author[l]{M.I.~Lopes,} 
\author[r]{A.~Manalaysay,}  
\author[d,x]{R.L.~Mannino,}  
\author[h]{M.F.~Marzioni,}   
\author[i,j,s]{D.N.~McKinsey,} 
\author[m]{D.-M.~Mei,} 
\author[u]{J.~Mock,}
\author[t]{M.~Moongweluwan,} 
\author[r]{J.A.~Morad,} 
\author[h]{A.St.J.~Murphy,} 
\author[p]{C.~Nehrkorn,}
\author[p]{H.N.~Nelson,} 
\author[l]{F.~Neves,}   
\author[i,j,s]{K.~O'Sullivan,}
\author[i]{K.C.~Oliver-Mallory,}
\author[b,c,d]{K.J.~Palladino,} 
\author[i,j]{E.K.~Pease,} 
\author[q]{C.~Rhyne,}
\author[p,v]{S.~Shaw,} 
\author[a,c]{T.A.~Shutt,} 
\author[l,1]{C.~Silva,\note{Corresponding author}}   
\author[p]{M.~Solmaz,}  
\author[l]{V.N.~Solovov,} 
\author[s]{P.~Sorensen,} 
\author[e]{T.J.~Sumner,} 
\author[u]{M.~Szydagis,}   
\author[n]{D.J.~Taylor,} 
\author[q]{W.C.~Taylor,} 
\author[j]{B.P.~Tennyson,} 
\author[x]{P.A.~Terman,} 
\author[f]{D.R.~Tiedt,}  
\author[b,c,y]{W.H.~To,} 
\author[r]{M.~Tripathi,} 
\author[i,j,s]{L.~Tvrznikova,}
\author[r]{S.~Uvarov,}   
\author[i]{V.~Velan,} 
\author[q]{J.R.~Verbus,} 
\author[x]{R.C.~Webb,} 
\author[x]{J.T.~White,}
\author[a,b,c]{T.J.~Whitis,} 
\author[s]{M.S.~Witherell,} 
\author[t]{F.L.H.~Wolfs,}  
\author[k]{J.~Xu,}
\author[e]{K.~Yazdani,}
\author[u]{S.K.~Young,} 
\author[m]{C.~Zhang.} 

\affiliation[a]{Case Western Reserve University, Department of Physics, 10900 Euclid Ave, Cleveland, OH 44106, USA} 
\affiliation[b]{SLAC National Accelerator Laboratory, 2575 Sand Hill Road, Menlo Park, CA 94205, USA} 
\affiliation[c]{Kavli Institute for Particle Astrophysics and Cosmology, Stanford University, 452 Lomita Mall, Stanford, CA 94309, USA}
\affiliation[d]{University of Wisconsin-Madison, Department of Physics, 1150 University Ave., Madison, WI 53706, USA}
\affiliation[e]{Imperial College London, High Energy Physics, Blackett Laboratory, London SW7 2BZ, United Kingdom}  
\affiliation[f]{South Dakota School of Mines and Technology, 501 East St Joseph St., Rapid City, SD 57701, USA} 
\affiliation[g]{University of Maryland, Department of Physics, College Park, MD 20742, USA} 
\affiliation[h]{SUPA, School of Physics and Astronomy, University of Edinburgh, Edinburgh EH9 3FD, United Kingdom}  
\affiliation[i]{University of California Berkeley, Department of Physics, Berkeley, CA 94720, USA} 
\affiliation[j]{Yale University, Department of Physics, 217 Prospect St., New Haven, CT 06511, USA} 
\affiliation[k]{Lawrence Livermore National Laboratory, 7000 East Ave., Livermore, CA 94551, USA}
\affiliation[l]{LIP-Coimbra, Department of Physics, University of Coimbra, Rua Larga, 3004-516 Coimbra, Portugal} 
 \affiliation[m]{University of South Dakota, Department of Physics, 414E Clark St., Vermillion, SD 57069, USA} 
\affiliation[n]{South Dakota Science and Technology Authority, Sanford Underground Research Facility, Lead, SD 57754, USA}
\affiliation[o]{Pennsylvania State University, Department of Physics, 104 Davey Lab, University Park, PA  16802-6300, USA} 
\affiliation[p]{University of California Santa Barbara, Department of Physics, Santa Barbara, CA 93106, USA} 
\affiliation[q]{Brown University, Department of Physics, 182 Hope St., Providence, RI 02912, USA} 
\affiliation[r]{University of California Davis, Department of Physics, One Shields Ave., Davis, CA 95616, USA}
\affiliation[s]{Lawrence Berkeley National Laboratory, 1 Cyclotron Rd., Berkeley, CA 94720, USA}  
\affiliation[t]{University of Rochester, Department of Physics and Astronomy, Rochester, NY 14627, USA}  
\affiliation[u]{University at Albany, State University of New York, Department of Physics, 1400 Washington Ave., Albany, NY 12222, USA} 
\affiliation[v]{Department of Physics and Astronomy, University College London, Gower Street, London WC1E 6BT, United Kingdom} 
\affiliation[w]{University of Massachusetts, Amherst Center for Fundamental Interactions and Department of Physics, Amherst, MA 01003-9337 USA} 
\affiliation[x]{Texas A \& M University, Department of Physics, College Station, TX 77843, USA}
\affiliation[y]{California State University Stanislaus, Department of Physics, 1 University Circle, Turlock, CA 95382, USA} 
\emailAdd{claudio@coimbra.lip.pt}
\date{\today}

\abstract{
The $(x, y)$ position reconstruction method used in the analysis of the complete exposure of the Large Underground Xenon (LUX) experiment is presented. The algorithm is based on a statistical test that makes use of an iterative method to recover the photomultiplier tube (PMT) light response directly from the calibration data. The light response functions make use of a two dimensional functional form to account for the photons reflected on the inner walls of the detector. To increase the resolution for small pulses, a photon counting technique was employed to describe the response of the PMTs. The reconstruction was assessed with calibration data including  ${}^{\mathrm{83m}}$Kr  (releasing a total energy of 41.5~keV) and ${}^{3}$H ($\upbeta^-$ with Q~=~18.6~keV) decays, and a deuterium-deuterium (D-D) neutron beam (2.45~MeV). Within the detector's fiducial volume, the reconstruction has achieved an $(x, y)$ position uncertainty of $\sigma$=~0.82~cm and $\sigma$~=~0.17~cm for events of only 200 and 4,000 detected electroluminescence photons respectively. Such signals are associated with electron recoils of energies $\sim$0.25~keV and $\sim$10~keV, respectively. The reconstructed position of the smallest events with a single electron emitted from the liquid surface (22~detected photons) has a horizontal $(x, y)$ uncertainty of 2.13~cm.}



\maketitle

\section{Introduction}

The LUX detector, a 370 kg liquid-gas dual-phase xenon time projection chamber (TPC) \cite{LUX_2013NIM}, has as its main scientific goal the observation of nuclear recoils resulting from hypothetical dark matter particle candidates, called Weakly Interacting Massive Particles. The excited xenon atoms and ionization electrons from the recoil in the liquid phase are observed through two different light signals: the prompt scintillation signal (called S1) and the charge signal (called S2). S1 arises from the direct production of xenon excited states and the recombination of some electrons with ions in the liquid. S2 is generated by the drifting of the ionization electrons that do not recombine towards the liquid surface, where they are extracted and accelerated into the gas phase producing electroluminesce. The light from both S1 and S2  is observed by two arrays of photomultiplier tubes (PMTs) placed on the top and bottom of the detector. The detection of both signals ensures a good reconstruction of the position of the interaction: the depth of the interaction is obtained from the time separation between  S1 and S2, while the $(x, y)$ coordinates are obtained from the distribution of the S2 light among the PMTs.  Assuming a parallel field geometry, the $(x, y)$ position of the interaction can be taken as that of the electroluminescence production (i.e.\ the S2 signal). The energy depositions of the recoils in the liquid xenon are considered point-like in the energy range of interest. For example, a 10~keV electron has a range of about 2~$\mu$m while that of a 5~MeV alpha particle is less than 50~$\mu$m. These numbers are much smaller than the typical size of the electron cloud in the $(x,y)$ plane due to electron diffusion along the transverse direction. At the center of the LUX chamber (drift time~=~160~$\upmu$s), the standard deviation of the electron transverse diffusion in the $x$ and $y$ direction is between 1~mm (for a diffusion constant of 30~cm$^2/$s \cite{Chen2012}) and 1.4~mm (for a diffusion constant of 60~cm$^2/$s \cite{PhysRevC.95.025502}).

A good 3D position reconstruction of the events is paramount in the analysis of dual phase detectors to ensure a good description of the observed events. Most of the radiological background events are localized closer to the walls, the photomultipliers or the electrode grids, leaving the center of the detector with a very low background \cite{LUX2015_BackgroundPaper}. Some especially troublesome backgrounds originate from the radioactive decay of ${}^{210}$Pb, ${}^{210}$Bi and ${}^{210}$Po plated on the Polytetrafluoroethylene (PTFE) walls of the detector (pp.\ 127--133, \cite{ChangLeePhDThesis}). These decays mimic low energy nuclear recoils and cannot be discriminated from a true recoil of a xenon nucleus using the ratio between S2 and S1.  We reject those events by preferably searching for signals in the detector central volume --- the fiducial volume. Consequently, a precise and accurate event position reconstruction, especially close to the walls of the detector and to  small S2 sizes (S2~=~0--4,000~detected photons, phd hereafter), is essential to accurately model external backgrounds from high rates at the walls to progressively lower rates towards the center of the detector.

Position reconstruction is also essential to correct for variations of the number of S1 and S2 photons throughout the chamber. These variations  are due to the light collection dependence on the interaction position, the finite electron lifetime due to the presence of impurities, and to local distortions of the electric field \cite{Fields2017}. The detector calibration with respect to nuclear recoils also requires a precise reconstruction of the vertex of neutron interactions in the detector \cite{LUX2016_DDCalibrations}.

Several $(x, y)$ position reconstruction algorithms have been employed in scintillation detectors. The oldest is the center of gravity or centroid method (often used in the Anger camera \cite{Anger1958_ScintillationCamera}), in which the position of the interaction is obtained from a weighted average of the PMT responses. A simple weighted average is biased towards the center of the array due to the finite extent, but the position can be partially corrected using a lookup table with corrections that depend on the position or an alternative parameterization \cite{Short_1984}. Another choice of position algorithm involves the use of artificial neural networks \cite{Morozov2016}, which have been employed by the XENON collaboration \cite{Pelssers_2015MT}.

Other  methods make use of a statistical test, such as the $\chi^2$ test or other maximum likelihood techniques \cite{Gray1976}. They determine the position of interactions by matching expected PMT outputs with the observed values. These methods have the advantage of giving an estimator (e.g.\ $\chi^2$ minimum) that can be used to assess the quality of the position reconstruction. For each PMT, the expected outputs as a function of the event position are usually stored in a Monte-Carlo generated lookup table (e.g.\ ref.\ \cite{Lindote2007}) or described by an empirical light response function (LRF), which characterizes the response of the PMT as a function of the position of the emission of the light $(x, y)$.

All the aforementioned algorithms require the use of some sort of calibration data for which the position of interaction is already known. In large detectors such as LUX, it is not feasible to use external radioactive sources to produce energy depositions at desirable locations in the inner region of the detector, due to the large volume and self-shielding properties of liquid xenon. Thus, the solution usually adopted is to make use of simulated data, which has limitations since the simulations may not describe accurately all the intricacies of the light collection such as the precise optical properties of the materials in these detectors. 

The position reconstruction algorithm employed in the LUX experiment (named Mercury, originally developed for the ZEPLIN-III dark matter experiment \cite{Solovov2011_PositionReconstruction}) is a statistical-based algorithm using a maximum likelihood test to find the best set of output parameters. Mercury uses S2 photons to reconstruct the $(x, y)$ position of the event. It employs LRFs to predict the response of each PMT for interactions at an arbitrary distance from that PMT. The innovative aspect and major advantage of this method is the way in which the LRFs are obtained: it employs a \emph{virtual scan} method, using the detector's own data with minimal or no reliance on simulations. In this virtual scan, the LRFs are obtained iteratively through a sequence of fits to the calibration data until the response functions converge simultaneously. This method is described in detail in \cref{section_The_Light_Response_Functions}.

LUX was designed to achieve a very low energy threshold (down to 1~keV for a nuclear recoil) aiming to be very sensitive to low mass WIMPs. This means that the signals of interest can be of very small amplitude (a few detected photons per PMT) which imposes new challenges to the position reconstruction. To optimize the precision and accuracy of the algorithm for this energy region, a new method was implemented (\cref{section_the_minimization_method}) to count individual photons in channels with very low signal. Another major difference from the original version of Mercury is that the LRFs in this work are 2-dimensional (axial and polar) instead of 1-dimensional to account for the reflection on the walls of the detector (\cref{section_The_Light_Response_Functions}). The ZEPLIN-III detector had low-reflectance internal walls, and thus the axial symmetry was a good approximation.

The position reconstruction method presented here was used in the analysis of the LUX data from 2013 (WS2013), including both the original analysis (85.3 live days of data, \cite{LUX2014_OriginalResults}) and the reanalysis (95.0 live days, \cite{LUX2015_ReanalysisPRL, LUX2016SpinDependent, LUX2017_Axions}), as well as in the analysis of the data collected from 2014 until 2016 (WS2014--16, 332.0 live days, \cite{LUX2016_SSR, LUX2017_SD}) and the calibration data used to monitor and calibrate the detector \cite{hertel2015, TritiumPaper2015, LUX2016_DDCalibrations}.
   
Mercury has also been used with some modifications in the DarkSide-50 experiment \cite{Darkside2015_FirstResults}, the Panda-X experiment \cite{PandaX_2016}, and proven to work well for a gamma camera for medical imaging \cite{Morozov2015_PosRec}.

This article is organized as follows: the method of obtaining the LRFs (virtual scan) is described in \cref{LRF_determination}; in \cref{section_the_minimization_method}, we introduce the method implemented to find the position of emission of S2 light; in  \cref{section_krypton_results}, we discuss the use of the position reconstruction algorithm on calibration data, the associated uncertainties, and the position resolution. 

\section{The Light response functions\label{section_The_Light_Response_Functions}\label{section_position_uncertainties}\label{LRF_determination}}

The determination of the shape of the light response functions (LRFs) is an essential feature in Mercury as they are used to estimate the expected response of the detector. The light response function of a PMT $i$, ${\mathcal H}_i$, is proportional to the probability that a S2 photon emitted at $(x,y)$ and detected by any of the PMTs is detected in channel $i$. These functions may change from PMT to PMT because the light collection depends on the relative position of the PMT in the array.  Besides the $(x, y)$ position, the minimization method described in   \cref{section_the_minimization_method} also estimates the associated uncertainties  and the value of the log-likelihood ratio, $q_{\mathrm{min}}$, associated to the reconstructed position.

In this section, we start with the description of the LUX detector geometry followed by the discussion of the
influence of reflections from the PTFE walls on the S2 light collection. Finally we describe the fitting procedure used to derive the LRFs.

\subsection{The LUX detector and calibrations\label{section_LUX_detector}}

\begin{figure}
 \begin{center}
  \includegraphics[width=7.0cm]{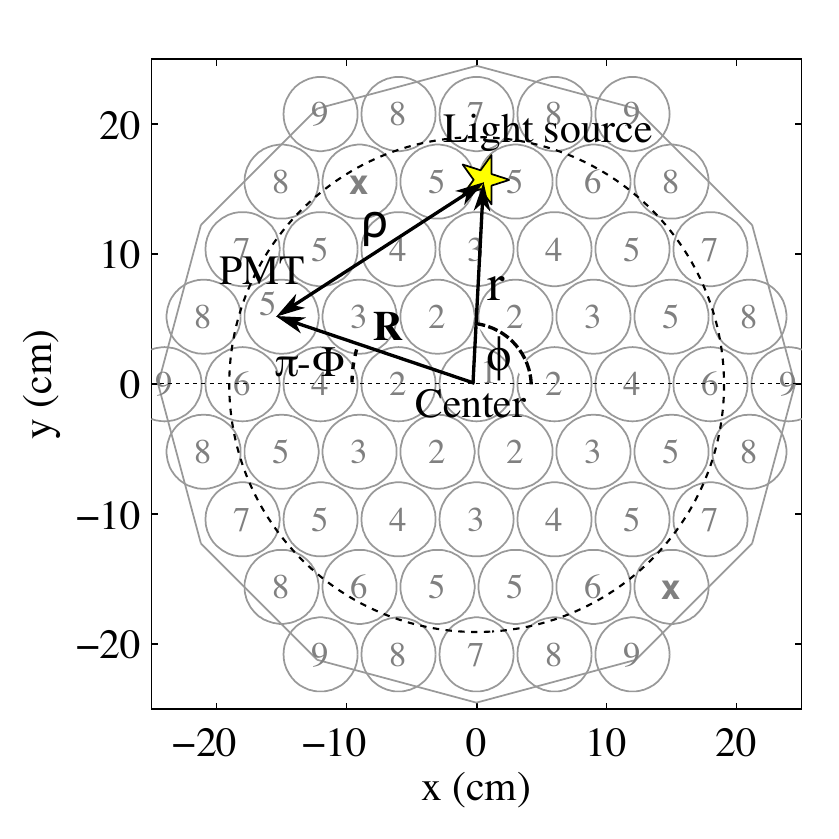}
  \caption{Top array of the LUX detector with the variables used in the definition of the functional form of the LRFs. The yellow star corresponds to the point of emission of the light from an electro-luminescence signal (S2). $\rho$ is the distance between the PMT and the light source;  $(R, \Phi)$ and $(r, \phi)$ are coordinates of the PMT and of the emission of the light position, respectively.  The PMTs are grouped according to the distance to the center of the chamber. The numbering scheme reflects the group which the PMT belongs to.
  }
  \label{Fig01_DefinicaoDasVariaveis}
 \end{center}
\end{figure}

The detector light collection, which determines the $(x, y)$ dependence of the LRFs, depends on many factors such as the geometry of the detector, the optical properties of the internal surfaces, and the PMT characteristics. In LUX, the active liquid xenon region has a dodecagonal shape with a height of 52.4~cm (with a 5.5~cm gas phase layer on top of the liquid), and a distance between the center of the side faces and the center of the detector of 23.65$\pm$0.05~cm \cite{LUX_2013NIM}. Both S1 and S2 signal result from far ultraviolet photons with a wavelength of $\sim$175~nm \cite{Jortner1965, FUJII2015293}. This light is detected by 122 Hamamatsu R8778 PMTs organized in two hexagonal arrays of 61~PMTs each, one placed immediately above and the other at the bottom of the sensitive volume \cite{LUX2013_PMTs}. The geometry of the top array is shown in  \cref{Fig01_DefinicaoDasVariaveis}. Each PMT has a diameter of 57 mm and an active photocathode diameter of 45~mm, except for the PMTs on the periphery which are partially covered by the detector walls. 

Each  of the 12 side walls of the detector and the space between the photomultipliers are covered by PTFE. The high reflectance of PTFE for the xenon scintillation light (>97\%  in the liquid phase \cite{FranciscoNeves2017_Reflectancia}) ensures a good light collection for both the S1 and S2 signals.

 During the WIMP search run, the detector was calibrated using a variety of radioactive sources, both internal (${}^{\rm 83m}$Kr and ${}^{3}$H, injected in the gas system of the detector) and external (2.45~MeV deuterium-deuterium (D-D) neutron source, ${}^{252}$Cf,  Am-Be, and ${}^{137}$Cs) \cite{LUX2015_ReanalysisPRD}. These calibrations were not only used to study the response of the detector for both electronic and nuclear recoils as a function of the deposited energy, but also to develop and study the position reconstruction method. The ${}^{\rm 83m}$Kr calibration is particularly important for the latter. It provided the data to i) obtain the LRFs of the PMTs (\cref{subsection_LRFDetermination}), ii) monitor the quality of the position reconstruction, and iii) measure the position resolution (\cref{section_krypton_calibrations}). 
 
 The ${}^{\rm 83m}$Kr isotope decays with a half life of 1.83\,h occurring in two transitions of 32.1~keV and 9.4~keV respectively, the half-life of the intermediate state being 154\,ns \cite{hertel2015}. Given the short time between these two decays, the two pulses overlap producing a single S2 signal that has between 4,000 and 20,000~phd distributed among several PMTs, depending on the depth of the event, which gives a sizable signal on the top PMT array but far from the saturation point ($\sim$10,000~phd per PMT). After being injected in the circulation system, ${}^{\rm 83m}$Kr distributes itself uniformly throughout the sensitive volume of the detector \cite{PhysRevC.80.045809}. The uniformity of the reconstructed ${}^{\rm 83m}$Kr events gives us insight about the quality of the position reconstruction. Furthermore, the large number of ${}^{\rm 83m}$Kr-decay events collected during the science runs (10 million events in WS2013) is sufficient to characterize with great detail the S2 light collection dependence on the event position. Single electrons are also a possible source of calibration. They are the lowest possible S2 signal ($\left<{\mathrm{S2}}\right>\sim$22~phd),  and thus can be used to study the systematics and statistical errors affecting the reconstruction of the lowest pulses which are of great interest in the LUX analysis.

The $(x, y)$ position reconstruction makes use of the signals from the PMTs of the top array except for two malfunctioning units (marked by a cross in \cref{Fig01_DefinicaoDasVariaveis}). Information from the bottom array PMTs is only used for the determination of the total S2 size, as the S2 light collection efficiency in each individual bottom PMT is almost independent on the $(x,y)$ position of emission in this high-reflectance chamber design. As for the S1 signal, its amplitude is relatively small on both PMT arrays, being inadequate to reconstruct the position. From the S2 signal of each PMT $i$ on the top array, two quantities are extracted for the $(x, y)$ position reconstruction:  the pulse area, $\mathcal{A}_i$, and the photon counts, $\mathcal{N}_i$. The pulse area, $\mathcal{A}_i$, is obtained by the software integration of the PMT signal along the duration of the pulse, while the number of detected photons, $\mathcal{N}_i$, is estimated by identifying and counting the individual discrete single detected photon pulses observed during the emission of the pulse (typically some microseconds for an S2 signal). The PMT signals are recorded in a dedicated Struck board with a sampling period of 10~ns \cite{LUXDAQ2012}, making it possible to implement photon counting  software. In the simple method employed, a photon is counted each time the photomultiplier waveform crosses a certain threshold (>1.4~mV, 5$\sigma$ above the baseline noise). Both $\mathcal{A}_i$ and $\mathcal{N}_i$ are used in the statistical method described in the section \ref{section_the_minimization_method}.

\subsection{Influence of the wall reflection in the S2 light collection}

In LUX, the LRFs are described using two-dimensional analytic functions, dependent on the radial position of the S2 light emission, $r$, and the distance between the PMT and the and the position of the S2 light emission, $\rho$. This is unlike the simpler approach in ZEPLIN-III described in \cite{Solovov2011_PositionReconstruction}, in which one dimensional LRFs were used instead as lateral reflections could be ignored to good approximation due to the low reflectivity of copper (the bi-hemispherical reflectance \cite{nicodemus} of polished copper in xenon gas and for $\lambda\sim$175~nm is less than 13\% \cite{Claudio2007}). In LUX the active region is defined by highly reflective PTFE panels extending up until the PMTs in the top array in order to increase the light collection of the S1 signal and thus increase the detection efficiency for low energy nuclear recoils. The bi-hemispherical reflectance of these panels in contact with gas was found to be ~70\% as measured by \cite{silva:064902} and >75\% as determined in LUX from comparison with Monte Carlo simulations \cite{LUX2015_ReanalysisPRD}. The bi-hemispherical reflectance of the panels immersed in liquid xenon is much higher (>95\% \cite{FranciscoNeves2017_Reflectancia}). While the light reflected on the PTFE walls in the gas gap has a significant impact in the light distribution pattern, the reflection of the scintillation light on the walls in contact with the liquid xenon does not influence the shape of the LRFs.

\begin{figure}
 \begin{center}
\includegraphics{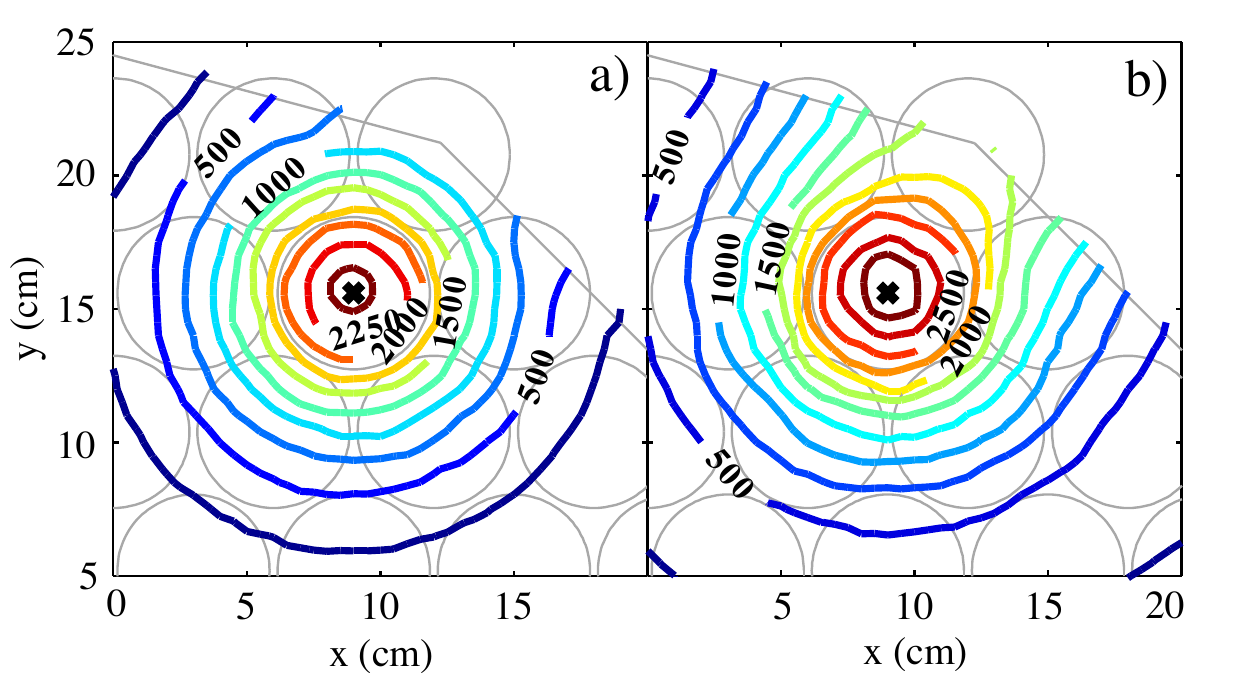}
\caption{Contour curves of the position of an interaction showing the number of photons detected by a PMT at $R$~=~18~cm  for non-reflective walls (a) and  100\% reflective PTFE panels (b). The data is from simulated ${}^{\mathrm{83m}}$Kr S2 events ($\sim$20,000 detected photons) uniformly distributed in the liquid xenon active volume.
The circular edge of the PMT borders and the border of the chamber are shown in gray.}

\label{Fig03_A_DensidadeXYReflective}
 \end{center}
\end{figure}

We studied the influence of the reflectance of the inner surfaces on the light collection distribution at each PMT of the top array using the S2s from a LUXSim ${}^{\mathrm{83m}}$Kr simulation with full light propagation \cite{LUX2012_LUXSim}. Two scenarios were considered: i) all the PTFE inner surfaces are diffusely reflective with a diffuse albedo of 100\%, and ii) all the PTFE inner surfaces are non-reflective. In both cases, specular reflection was not considered. For each case, 20,000 ${}^{\mathrm{83m}}$Kr events uniformly distributed in the active liquid volume were generated, and the light was propagated until it was detected or absorbed. \Cref{Fig03_A_DensidadeXYReflective} shows the S2 light collected at a PMT close to a detector wall. As shown, for the case with non-reflective inner surfaces, the PMTs collect almost only direct S2 light, which is emitted isotropically, with exception of two small components: light that is reflected on the liquid/gas interface and light that crosses that interface and returns to the gas via Rayleigh scattering in the liquid. On the contrary, for reflective surfaces, the light collection is no longer axially symmetric, showing a saddle extending from the center of the photomultiplier towards the wall. It is clear from this study that in LUX the light collection  depends not only on the distance between the PMT and the S2 position but also on the distance of the S2 position to the  walls, requiring two variables to describe the LRFs.

\subsection{LRF determination\label{subsection_LRFDetermination}}

We implemented a new model for the LRFs that takes into account the light reflected on the walls and takes advantage of the symmetries of the detector by adding an axial ($\eta$) and a polar ($\varepsilon$) component
calculated for each PMT $i$ on the top array. This function can be written as
\begin{equation}
 {\mathcal H}_i\left(r, \rho \right) = {\mathcal C}_i \left[\eta\left(\rho\right) + \varepsilon_i\left(r, \rho \right)\right],
 \label{response_function}
\end{equation}
where $\rho$ corresponds to the distance between PMT $i$ centered at ($R_i$, $\Phi_i$) and the light emission at ($r$, $\phi$) (\cref{Fig01_DefinicaoDasVariaveis}).  ${\mathcal C_i}$ are normalization constants to equalize the response of the PMTs for the same number of incident photons per unit of PMT window area under the same geometrical conditions. Differences in the values of ${\mathcal C_i}$ are due to different quantum efficiencies of the photomultipliers for xenon scintillation and different exposed photocathode areas as some PMTs are partially covered with PTFE tiles reducing the light collection.

The functional form of the axial component was  determined using LUXSim \cite{LUX2012_LUXSim}, being described by the following empirical function
\begin{equation}
 \eta\left(\rho\right) = \frac{A}{\left(1+\gamma^2 \rho^2\right)^{\frac{3}{2}}} + m\rho+b,
\label{eq_radial_component}
\end{equation}
where $A$, $\gamma$, $m$ and $b$ are the fitting parameters. The first term, a bivariate-Cauchy function, describes the light that goes directly to the PMT or is reflected on the gas/liquid interface, while the polynomial term is necessary to describe the light that is reflected inside the liquid bulk or in the PTFE trefoils placed between the PMTs. 

The polar component, $\varepsilon$, cannot be described using the same function for all PMTs, as in the case of $\eta$, given that PMTs closer to the walls detect more reflected light. However, the symmetry of the LUX detector allows us to use the same LRF for PMTs with the same radial position $R_i$. Therefore, the function $\varepsilon$ has to be found for each of the 8 different groups of PMTs in the top array (identified in \cref{Fig01_DefinicaoDasVariaveis} with the numbers 2--9). This approach significantly reduces the number of parameters and is more robust as the symmetry of the chamber is directly incorporated in the LRF model. In the initial iterations, only the PMT groups closer to the walls ($R_i$~>~19~cm), where the polar component is more significant, are used (the border of $R_i$~=~19~cm is represented in \cref{Fig01_DefinicaoDasVariaveis} by the dashed line). After at least 5 iterations, all the PMT groups are incorporated in this fit. 

The shape of the curve used to fit the polar component to the experimental data was also inferred from simulations of the light collection in the detector. These simulations showed that this component is described in first approximation by the following empirical function:
\begin{equation}
\varepsilon_g \left(\rho, r\right) = \varkappa_g \exp{\left(\frac{r}{\xi_g}\right)}\exp{\left(-\frac{\rho}{\zeta_g}\right)},
\label{eq_polar_component}
\end{equation}
where the index $g$ indicates the PMT group ($g$~=~2--9) and $\varkappa_g$, $\zeta_g$ and $\xi_g$ are the fit parameters. This function ensures that the polar component increases with the radial coordinate of the light emission location, $r$, as expected, since the effect of the wall reflection is larger for  events at larger radii. However, it was clear that this simple function was not sufficient to describe the intricacies of the light collection in the detector. Hence, in the final iterations of the process of the LRF determination that will be described next, an alternate format of \cref{eq_polar_component} that included more fitting parameters was used:
\begin{eqnarray}
\varepsilon_g \left(\rho, r\right)  & = & \left[\varkappa_g \exp{\left(\frac{r}{\xi_g}\right)} + \alpha_g(r) \right]\exp{\left(-\frac{\rho}{\zeta_g + \beta_g(r)}\right)}\nonumber\\
\alpha_g\left(r\right)  & = & k_{g,1} \exp{\left[-\frac{\left(r-k_{g,2}\right)^2}{2\left(k_{g,3}\right)^2}\right]} + k_{g,4} r + k_{g,5} \nonumber \\
\beta_g\left(r\right)  & = & k_{g,6}\exp{\left[ -\frac{\left(W_c-r\right)^2}{2\left(k_{g,7}\right)^2}\right]},
\label{eq_polst}
\end{eqnarray}
where $\left(\varkappa_g, \xi_g, \zeta_g, k_{g,1}, k_{g,2} ...... k_{g, 7}\right)$ correspond to the fitting parameters for each group of PMTs and $W_c$~=~24.5~cm is the position of the corners of the wall in the LUX dodecagon. Although this function is more complex than \cref{eq_polar_component}, the new terms only introduce small corrections to the LRFs. $\varkappa_g$ is larger than $\alpha_g$, and the constant term $\zeta_g$ dominates with an average value  $\sim$0.19\,cm, while $k_{g,6}$ is only, on average, $\sim$0.04\,cm.

\Cref{response_function} is used for all the PMTs with the exception of the central channel ($R$~=~0). As this PMT has axial symmetry relative to the active volume of the chamber, its LRF can be described using the axial component, $\eta(\rho)$, only (given by \cref{eq_radial_component} but in an independent fit).

\begin{figure}
 \begin{center}
  \includegraphics{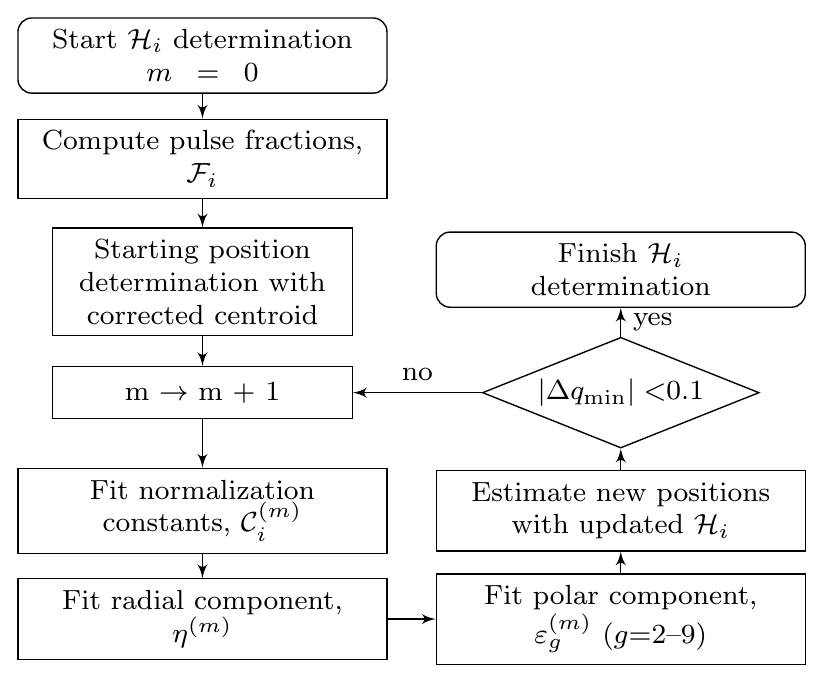}
  \caption{Flowchart of the LRF fitting process.}
  \label{Figure05_Flowchart}
 \end{center}
\end{figure}

The experimental data used to obtain the LRFs consisted of a sample of ${}^{{\rm 83m}}$Kr calibration S2 pulses with unknown positions. For this, we selected 100,000 ${}^{{\rm 83m}}$Kr events close to the top of the detector (at depths ranging from $\sim$8~cm to $\sim$12~cm). This  number of events ensures the fitting is done in a reasonable time and is sufficient to describe the light collection in the detector. For each ${}^{{\rm 83m}}$Kr event and for each PMT, the pulse fractions, $\mathcal{F}_i$, are defined as 
\begin{equation}
\mathcal{F}_i =\mathcal{A}_i {\Large /} \mathcal{A}_T  \quad \textrm{or} \quad \mathcal{F}_i =\mathcal{N}_i {\Large /} \mathcal{A}_T,
\label{the_second_eq}
\end{equation}
$\mathcal{F}_i$ being a sample of the function $\mathcal{H}_i\left(r,\rho\right)$. ${{\mathcal A}}_T$ is the total pulse area of the S2 pulse summed over all the photomultipliers in the chamber.

The functions $\mathcal{H}_i\left(r, \rho\right)$ are obtained by an iterative process. The flowchart of the fitting process that follows is shown in \cref{Figure05_Flowchart}. In each iteration $m$, $\mathcal{C}_i^{(m)}$, $\eta^{(m)}\left(\rho\right)$, and $\varepsilon_i^{(m)}\left(r, \rho\right)$ are fitted in this order. The functional forms of $\eta\left(\rho \right)$ and $\varepsilon_i\left(r, \rho\right)$ are fitted to their values calculated with \cref{response_function} using the experimental values of $\mathcal{H}_i$ given by $\mathcal{F}_i$ and the event position estimates from the previous iteration. The constants $\mathcal{C}_i^{(m)}$ and the functions $\eta^{(m)}$ and $\varepsilon_i^{(m)}$ are then employed to get better estimates of the positions of the experimental events using the statistical method described in \cref{section_the_minimization_method}. These positions are used in the next iteration. The event positions used in the first iteration are obtained with a corrected centroid method \cite{Solovov2011_PositionReconstruction}. 

The constants $\mathcal{C}_i^{(m)}$ are obtained by equalizing the radial response of the PMTs from the top array. The method used to determine $\mathcal{C}_i^{(m)}$ is as follows: the observed axial component, $\eta_{\mathrm {obs}}$,  is computed as $\eta_{\mathrm {obs}} = \mathcal{F}_i/\mathcal{C}_i^{(m-1)} - \varepsilon_i^{(m-1)}$ (in the first iteration $\mathcal{C}_i^{(m-1)}$ and $\mathcal{\varepsilon}_i^{(m-1)}$ are set equal to 1 and 0, respectively). The values of $\eta_{\mathrm {obs}}$ with $\rho\in[0,15]$~cm are grouped in sections 1~cm wide in $\rho$. For each section, we calculate the ratio between the average over that section of the axial function for PMT $i$ and that for all the PMTs of the top array. The factor $\mathcal{C}_i^{(m)}$ is given by the average over all the sections of these ratios multiplied by $\mathcal{C}_i^{(m-1)}$. 

To obtain $\eta^{(m)}\left(\rho\right)$, the function given by \cref{eq_radial_component} is fitted to its values at the experimental { values} given by $\mathcal{F}_i/\mathcal{C}_i^{(m)} - \varepsilon_i^{(m-1)}$ (in the first iteration $\varepsilon_i^{(m-1)}$ is set equal to zero). 

For the determination of both $\mathcal{C}_i^{(m)}$ and $\mathcal{\eta}^{(m)}$, only events within a strip with thickness $d$ along the radial line between the current PMT center and the farthest wall are used, selected by the conditions
\begin{equation}
\left|r\sin\left(\phi - \Phi_i\right)\right| < d \quad {\rm{and}} \quad   r\cos\left(\phi - \Phi_i\right) < R_i.
\label{SelectionCondition}
\end{equation}
In this fit, we set $d$~=~0.25~cm.

To determine $\varepsilon_i^{(m)}$, the function given by \cref{eq_polar_component} (or by eq.\ (\ref{eq_polst}) for the final iterations) is fitted to its values at the experimental event positions given by $\mathcal{F}_i/\mathcal{C}_i^{(m)} - \eta^{(m)}$. 

\begin{figure}
 \begin{center}
  \includegraphics[width=8.5cm]{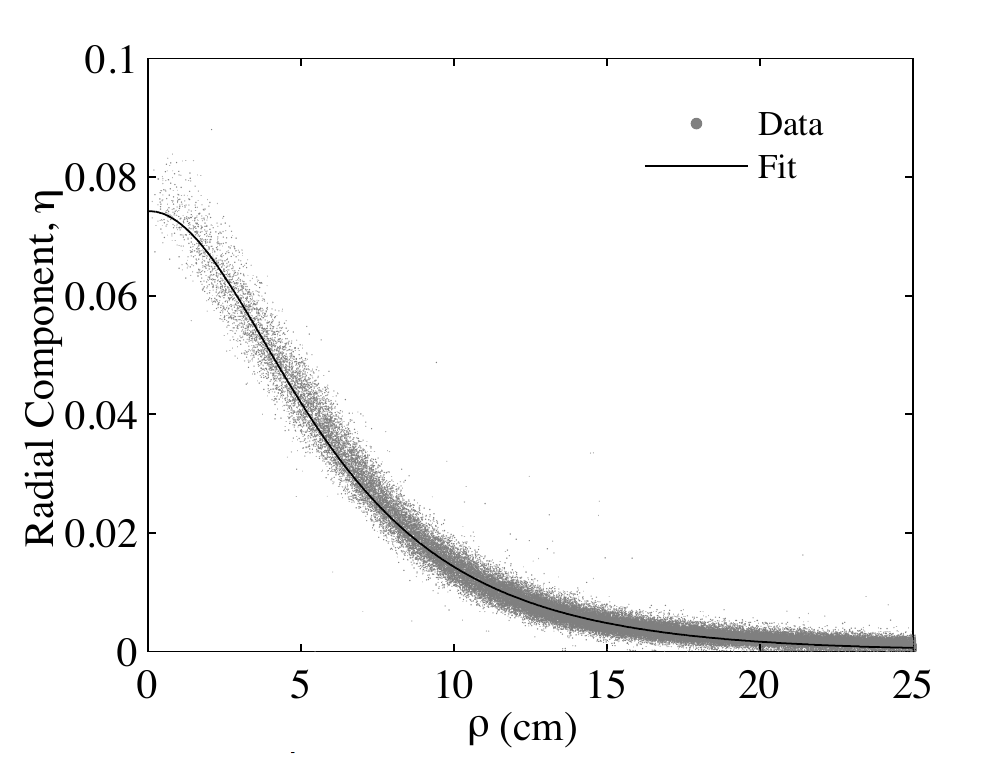}
  \caption{The axial component of the LUX PMTs. $\left\{\mathcal{F}_i/\mathcal{C}_i - \varepsilon_i\right\}$  is shown as function of $\rho$ --- the distance between the center of the PMTs and the position of the event. The polar component as predicted by the function $\varepsilon$ is being subtracted. The curve is the fit with  \cref{eq_radial_component}.}
  \label{Fig06_RadialComponent_Data}
 \end{center}
\end{figure}

The $\mathcal{C}_i^{(m)}$, $\eta^{(m)}$, and $\varepsilon_i^{(m)}$ are then used to update the estimated coordinates of the positions of the experimental events (using the statistical method described in  \cref{section_the_minimization_method}) that will be used in the next iteration. In the first 5 iterations, a uniformization of the event positions (see section IV in ref.\ \cite{hertel2015}) obtained in the previous iteration was performed in the radial direction, in order to speed-up the process.

\begin{figure*}
 \begin{center}  
\includegraphics[width=\textwidth]{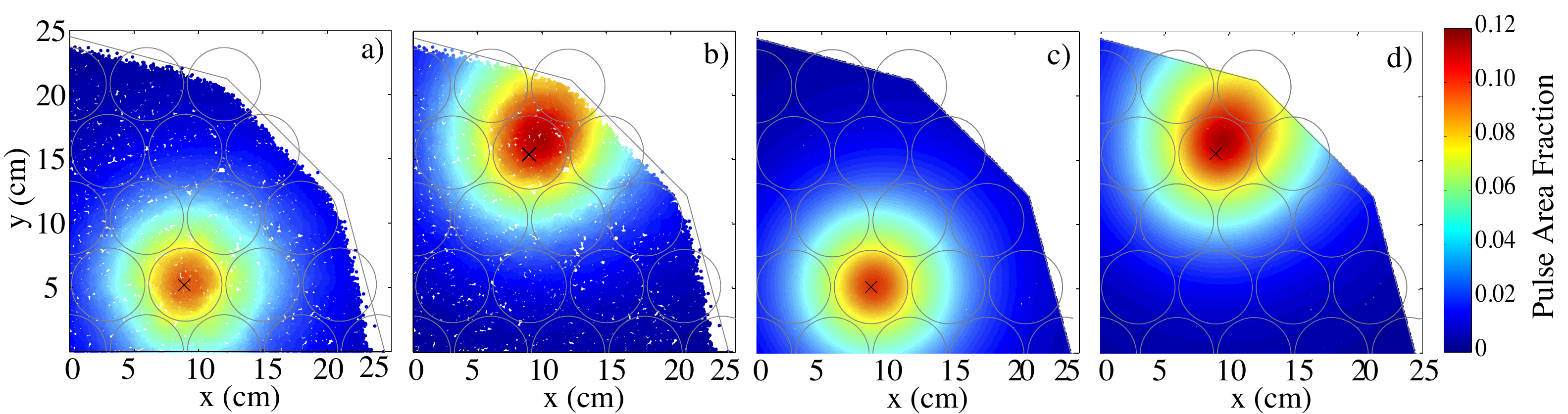}
\caption{2D histogram of the fractions ${\mathcal F}_i$ from ${}^{{\rm 83m}}$Kr for
$R_i$~=~10.4~cm (panel a) and $R_i$~=~18~cm (panel b)  events and the respective LRFs (panel c and d). The contours of the photomultipliers and the border of the chamber are shown in gray, while the center of the PMT under analysis is marked by a black cross.}
  \label{Fig07_Mercury_LRFs}
 \end{center}
\end{figure*}

The iteration process finishes once the average value of the $q_{\mathrm{min}}$ (minimum of the likelihood ratio defined next, \cref{LikelihoodRatio}) does not change significantly between consecutive iterations ($\left|\Delta \left<q_{\mathrm{min}}\right>\right|$~<~0.1 typically). For LUX, the number of iterations needed to find a good set of functions $\left\{\mathcal{H}_i\right\}$ was around 20, clearly larger than than the 5 iterations needed in ref.\ \cite{Solovov2011_PositionReconstruction} when the simpler axial approximation was used.

The LRFs of the faulty PMTs (marked by an x in the figure 1) are given by the average response of the PMTs placed at the same distance $R_i$.

The good agreement between the data and the LRFs { is illustrated in \cref{Fig06_RadialComponent_Data,,Fig07_Mercury_LRFs}}. The former shows the final iteration of the fit of \cref{eq_radial_component} to the experimental data; in \cref{Fig07_Mercury_LRFs}, the colors represent the observed  fractions $\mathcal{F}_i$ (panels a and b) or the corresponding LRFs (panels c and d) as a function of the reconstructed interaction position $(x, y)$ for a PMT in the central region ($R_i$~=~10.4~cm) and a PMT close to the walls of the detector ($R_i$~=~18~cm). The LRFs are mostly axial symmetric for  PMTs in the central region, approaching an oval shape for PMTs closer to the walls due a larger polar component. In the latter case, the position of the maximum is not coincident with the center of the PMT being pushed towards the wall of the detector as already expected from the simulation results shown in \cref{Fig03_A_DensidadeXYReflective}. 

We found that a single set of LRFs was sufficient to reconstruct all the collected data for the full detector exposure. In fact, we monitored the average value of $\left<q_{\mathrm{min}}\right>$ for the ${}^{\mathrm{83m}}$Kr along both runs. For both the WS2013 and WS2014--16 results, $\left<q_{\mathrm{min}}\right>$ was stable within 1\% with no visible degradation observed in the position reconstruction. Also, we fitted the LRFs using other krypton datasets from WS2013 and we found that they were mostly indistinguishable. Additionally, we monitored the PMT gain stability and the light yield stability.  It was found that the PMT gain fluctuations were smaller than 2\% and that the relative variation in the light yield was only about 0.6\% \cite{LUX2015_ReanalysisPRD}.

\section{The statistical method\label{section_the_minimization_method}}

A maximum likelihood algorithm \cite{Cowan1998, Macovski_1976} was implemented to search for the $(x, y)$ position of the interaction that would produce in an array of $N$ PMTs the expected outputs $\mathcal{S}=\{\mathcal{S}_i(x,y)\}$  ($i=1,...,N$) which are the closest to  the observed outputs $\mathcal{D}=\{\mathcal{D}_i\}$. This consists in finding the position $(x, y)$ that maximizes the   likelihood function,
\begin{equation}
\mathcal{L}\left( \mathcal{D}|\mathcal{S}  \right) = \prod_{i=1}^{N} p_i\left( \mathcal{D}_i|\mathcal{S}_i \right), \label{MLM_method}\\
\end{equation}
where $p_i$ is the probability for the PMT $i$ to have the observed output $\mathcal{D}_i$ given the expected output $\mathcal{S}_i$.  Provided that the response of the PMTs and the subsequent electronics is linear, the expected outputs can be written as 
\begin{equation}
\mathcal{S}_i = {\mathcal D}_T {\mathcal H}_i \left(x, y\right),\quad {\rm with} \quad {\mathcal D}_T = \sum_{i=1}^{\mathrm{all\,PMTs}} \mathcal{D}_i,\label{expected_outputs}
\end{equation}
where  ${{\mathcal D}}_T$ is the total number of detected photons summed over all the photomultipliers in the chamber. For the non-working PMTs, we assumed ${\mathcal D}_i={\mathcal S}_i$. The LUX analog signal chain and DAQ maintain linearity  up to energies of $\sim$100~$\rm keV_{ee}$ ($\mathcal{A}_{\rm top}\simeq$\,50,000 in the S2) which is well above the  WIMP region of interest and the $^{83\mathrm{m}}$Kr and ${}^{3}$H $\upbeta$ calibrations ~\cite{LUXDAQ2012}.

Instead of maximizing $\mathcal{L}\left( \mathcal{D} | \mathcal{S}\right)$, it is more convenient to minimize the  log-likelihood ratio \cite{logLikelihood} given by
\begin{equation}
{q} = -2\left[\ln\mathcal{L} - \ln\mathcal{L}_0\right] \quad
 \text{with} \quad \mathcal{L}_0 = \mathcal{L}\left(\mathcal{D}|\mathcal{D}\right),
 \label{LikelihoodRatio}
\end{equation}
where $\mathcal{L}_0$ corresponds to the likelihood maximized in an unconstrained way (in this analysis, we assumed $\mathcal{S}=\mathcal{D}$ in the computation of $\mathcal{L}_0$).

The probability $p_i$ can be written as
\begin{equation}
p_i\left(\mathcal{D}_i|\mathcal{S}_i\right)  =  \sum_{n=0}^{+\infty}P_i(n;\mathcal{S}_i) \cdot u_i(n;\mathcal{D}_i),
\end{equation}
where $P_i(n;\mathcal{S}_i)$ describes the fluctuations of the number of detected photons, $n$, for an expected output $\mathcal{S}_i$; and $u_i(n;\mathcal{D}_i)$ characterizes the response of the PMT $i$ and the respective signal processing electronics, being equal to the probability that the response is $\mathcal{D}_i$ for $n$ detected photons. In this work we assume $P_i(n;\mathcal{S}_i)$ to be the same for all the PMTs and given by a Poisson distribution:
\begin{equation}
P_i(n;\mathcal{S}_i) = \left[\mathcal{S}_i^n\exp(-\mathcal{S}_i)/n! \right].
\end{equation}

As previously mentioned, the PMT outputs $\mathcal{D}_i$ can be assessed by pulse areas or by photon counting. The formulation of the maximum likelihood method in each of these cases is described next, along with an innovative mixed method, which uses both as proxies for $\mathcal{D}_i$ depending on the pulse area recorded for each PMT.

\subsection{Pulse areas\label{subsection_pulse_areas}}

In the pulse area method, the PMT response for a single detected photon is  $u(1)$ with  $\left<u(1)\right>$~=~1~phd. Therefore, the response for $n$ detected photons ($n$~>~1) is obtained by the following recurrent convolution
\begin{equation}
	u(n) = u(n-1) \ast u(1).
  \label{multiplephotoelectronresponse}
\end{equation}

 The probability density $p_i$ is given by the sum of the response for $n$ detected photons with weights given by Poisson  distributions and taking the pulse areas, $\mathcal{A}=\{\mathcal{A}_i\}$, as estimators for the measured outputs, $\{\mathcal{D}_i\}$, leading to
\begin{equation}
p_i\left(\mathcal{A}_i| \mathcal{S}_i \right) = \mathrm{e}^{-\mathcal{S}_i}\left[\delta\left(\mathcal{A}_i\right) + \sum_{n=1}^{+\infty} \frac{\left(\mathcal{S}_i\right)^n}{n!} u_i\left(n; \mathcal{A}_i\right) \right],
 \label{probability_dis}
\end{equation}
where the Dirac delta function, $\delta\left(\mathcal{A}_i\right)$, corresponds to the case $n=0$, and the expected outputs, $\mathcal{S}_i$, are obtained from
\begin{equation}
\mathcal{S}_i = {\mathcal A}_T {\mathcal H}_i \left(x, y\right),\quad {\rm with} \quad {\mathcal A}_T = \sum_{i=1}^{\mathrm{All\,PMTs}} \mathcal{A}_i.
\end{equation}
To obtain the likelihood ratio $q$, we can use the probability distribution from \cref{probability_dis} in \cref{LikelihoodRatio}.

When the number of photons detected in each PMT is large enough, Gaussian statistics can be used for $P_i$ instead of the Poisson distribution. In the asymptotic Gaussian limit (Wilks' theorem \cite{Wilks1938_LikelihoodRatio}), $q$ is distributed as the following simple $\chi^2$ minimization 
\begin{equation}
q\equiv \chi^2 = \sum_{i=1}^N\frac{\left({\mathcal S}_i-\mathcal{A}_i\right)^2} {\mathcal{S}_i\left(1+\sigma_i^2\right)},
 \label{chi2_method}
\end{equation}
where the term $\left(1+\sigma_i^2\right)$ represents the excess noise factor for each PMT \cite{5402300}. This factor measures the degradation of the signal compared with Poissonian statistics. It can be shown that $\sigma_i$ is the standard deviation of the response to single  photons. 

The pure pulse area method was used in the initial analysis of the WS2013 data \cite{LUX2014_OriginalResults}. In this analysis, the single photon response, $u(1)$, was obtained from LED calibrations ($\lambda$~=~420\,nm). It was found that $u(1)$ was well described by a Gaussian  for all PMTs with the value of $\sigma_i$ ranging from 0.4 to 0.6~phd \cite{LUXCF_thesis}. The simpler $\chi^2$ minimization was used for  ${\mathcal A}_T>2,000$~phd. 

\subsection{Photon counting}

In the photon counting technique, the response of a PMT for $n$ detected photons is $u(n)=\delta(n)$; thus, $p_i$ follows a Poisson law, $q$ being given by (equation 39.16 in ref.\ \cite{PDG2014})
\begin{equation}
q  =  2\sum_{i=1}^N \left[{\mathcal S}_i-\mathcal{N}_i + \mathcal{N}_i\ln\frac{\mathcal{N}_i}{\mathcal{S}_i}\right].\\
\label{LikelihoodPoisson}
\end{equation}
For the PMTs with no photon detection ($\mathcal{N}_i=0$), we have $\mathcal{N}_i\cdot\ln(\mathcal{N}_i/\mathcal{S}_i)=0$.

\begin{figure}
 \begin{center}
  \includegraphics[width=8.5cm]{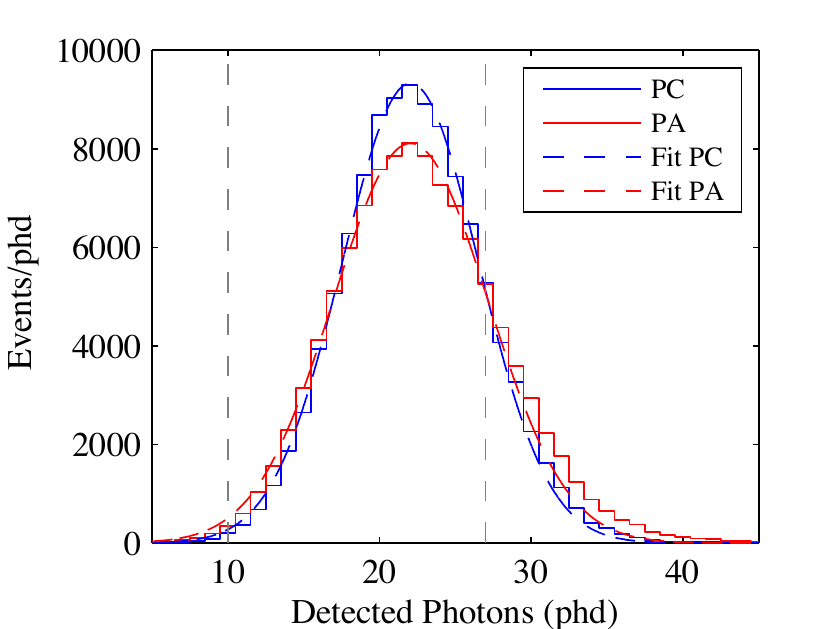}
  \caption{
  Distribution of the pulse areas (PA) and photon counts (PC) for pulses with a single extracted electron from the liquid. The single electrons were emitted between the S1 and the S2 of the ${}^{\rm 83m}$Kr. Both histograms are fitted with Gaussian distributions  for pulse size between 10 and 27~phd (fit-range indicated by the dashed vertical lines), the upper bound  defined to avoid contamination from double electron pulses. The fit to the pulse areas is centered at 22.05$\pm$0.09~phd with a resolution ($\sigma/\mu)$ of 23.1$\pm$0.6\%, while the fit to the photon counting histogram has a mean of 22.04$\pm$0.06~phd and a resolution of 20.5$\pm$0.4\%. This improvement of 2.6$\pm$0.7\% matches well the expected 2.7\% estimated from the reduction of the noise factor. Only events with a reconstructed radius smaller than 16~cm were considered in this analysis.}
 \label{Fig02_Mercury_PCVsPA}
 \end{center}
\end{figure}

The main advantages of using the counting over the pulse area method are that it is much less computationally  expensive, and the excess noise factor  is 1, which improves the energy resolution. Moreover, the pulse areas method has a poor performance for very low signals given that the fluctuations in the baseline noise are integrated in the signal, while in the photon counting method most of the baseline noise is well rejected by the threshold discriminator. The latter fact can be directly assessed from \cref{Fig02_Mercury_PCVsPA} which shows a  comparison of the resolution  between these two metrics using pulses from a single extracted electron.

Another main advantage of photon counting is the manner in which the double photoelectron emission from a single xenon scintillation photon is treated \cite{Faham2015}. For xenon scintillation light, 20\% of the photons produce two photoelectrons per photon on the photocathode of the PMT. In this case,  the single photon response function, $u(1)$, can no longer be approximated by the Gaussian single photon response of the PMTs obtained with blue LED light, in which case the photons do not have enough energy for double photoelectron production. This greatly complicates the use of the pulse area integration method that requires the knowledge of $u(1)$, but has no effect on the photon counting method, as the two photoelectrons are emitted simultaneously and thus counted as corresponding to a single photon.

Photon counting is limited by the pile-up that occurs when two or more photons are detected almost simultaneously, producing a signal indistinguishable from that of single photon. The probability of pile-up depends on the time resolution, $\tau$, assumed to be the same for all the photomultipliers, and on the photon flux. The time resolution, $\tau$, is the minimum time difference between two detected photons to ensure they are not merged into a single count. It was estimated from the analysis of S1 signals to be 30~ns.

To estimate the effect of pile-up, we approximated the light flux during the emission of the S2 signal to a Gaussian distribution defined by a standard deviation $\sigma_{\mathrm S2}$, estimated for each S2 pulse. $\sigma_{\mathrm S2}$ increases with the drift time of the event due to longitudinal diffusion of the electrons while drifting to the liquid surface \cite{Sorensen2011}. From the analysis of the S2 signals, it was found that the average value of $\sigma_{\mathrm S2}$ ranges from 360\,ns to 830\,ns for events from the top and bottom of the liquid layer, respectively. The probability of pile-up was computed in a dedicated Monte-Carlo simulation for different values of $\sigma_{\mathrm S2}$ and pulse areas of the PMTs. These results were then used in the $q$ of the mixed method described in the following subsection.

\subsection{Mixed method}

In the mixed method, the response of the PMTs is described with the photon counting method when the pulse area, ${\mathcal A_i}$, observed for a PMT $i$, is smaller than a certain threshold, $T$, and the pulse areas method above that threshold. The threshold depends on the pulse width of the event, and the log-likelihood ratio is given by the value of $q$ for the PMTs described by pulse areas plus the value of $q$ for the PMTs described by photon counting. By combining \cref{chi2_method,,LikelihoodPoisson} we have:
\begin{align} 
q & =  \sum_{\substack{i=1,\\ \mathcal{A}_i<T}}^{N} 2 \left[{\mathcal S}_i-\mathcal{N}_i + \mathcal{N}_i \ln\frac{\mathcal{N}_i}{\mathcal{S}_i}\right]
 + \sum_{\substack{i=1,\\\mathcal{A}_i>T}}^{N}\frac{\left({\mathcal S}_i-\mathcal{A}_i\right)^2}{\mathcal{S}_i\left(1+\sigma_i^2\right)}.
\label{eq_loglikelihood}
\end{align} 
The threshold, $T$, is set such that the contamination of the pile-up in the photon counting is smaller than 5\%, determined from the Monte-Carlo simulations described in the previous sections. It was found to be given by
\begin{equation}
T = 0.39\cdot \frac{\sigma_{\mathrm S2}}{\tau}~{\mathrm{(phd)}},
\end{equation}
where $\tau$~=~30~ns.

The mixed method was used both in the reanalysis of the WS2013 results \cite{LUX2015_ReanalysisPRL} and in the WS2014--16 results \cite{LUX2016_SSR}. In both analyses, the pulse areas are calibrated using ${}^{\rm 83m}$Kr calibration data in order to ensure compatibility between photon counting and pulse areas \cite{LUX2015_ReanalysisPRD}. The mean values of the two histograms on \cref{Fig02_Mercury_PCVsPA}  agree within  uncertainties, showing good agreement between the photon counting and the pulse areas methods.

\subsection{Position uncertainties\label{section_Likelihood_Ratio_Analysis}}

\begin{figure*}
 \begin{center}
     \includegraphics[width=\textwidth]{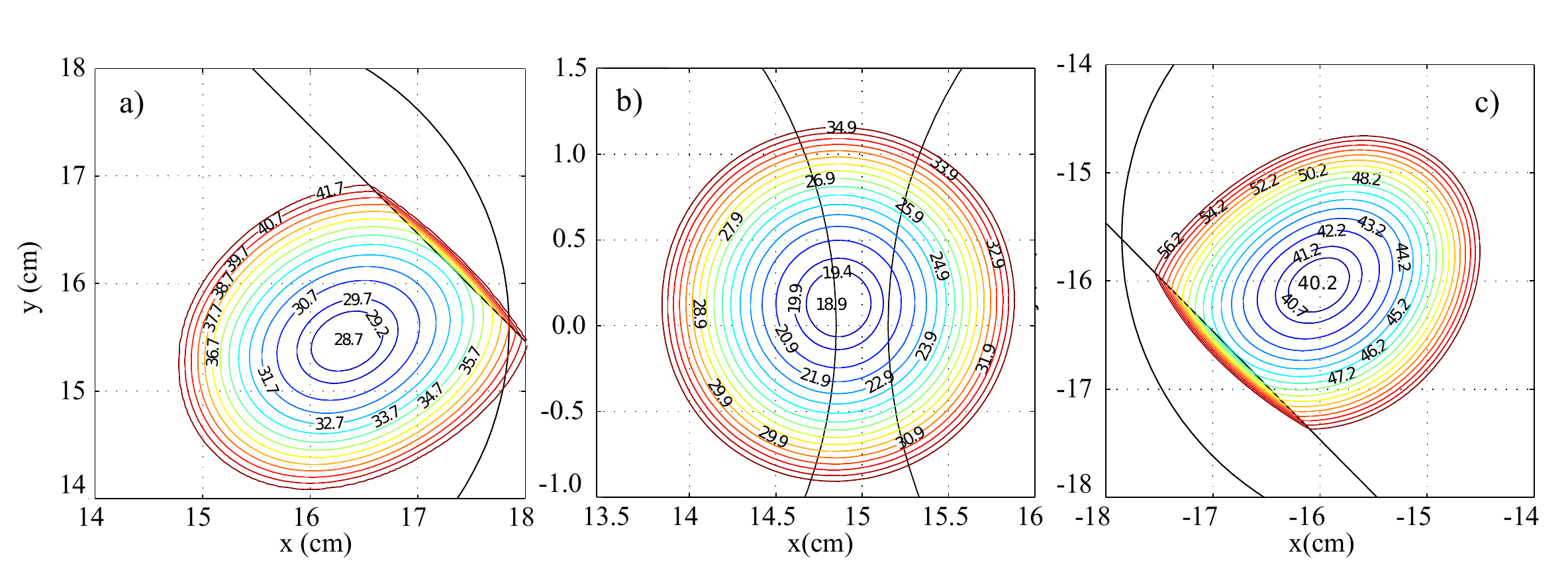}	
\caption{Contours of the $q$ minimization profile for three different small S2 pulses (${\mathcal A}_T$~<~4,000~phd): two near the walls of the detector (a and c) and the other close to the central region of the PMT array, between two PMTs. PMT shapes and the border of the chamber are represented  by black lines.}
  \label{Fig11_Chi2Curves}
 \end{center}%
\end{figure*}

The analysis of the shape of $q$ around the minimization point (${q}_{\mathrm{min}}$) is used to estimate the uncertainties associated with the position reconstruction. \Cref{Fig11_Chi2Curves} shows the contour lines of the $q$ minimization for three different small S2 events ($\mathcal{A}_T$<4,000~phd). As shown, the $q$ surface is smooth, well behaved and has an unambiguous global minimum. The $q$ is axially symmetric for events in the central region of the detector ($r$<20~cm), while for events near the walls the surface is elongated along the radial direction, reflecting the larger uncertainty on that coordinate.

Considering the symmetry of the $q$ contours, the  uncertainties in the reconstructed positions are best expressed in cylindrical coordinates ($\sigma_r$, $\sigma_\phi$). To obtain them, we select the curve in the $(x, y)$ space where  $q = q_{\rm min} + 1$ \cite{PDG2014}. The azimuthal, $\sigma_\phi$, and the radial, $\sigma_r$, uncertainties are given by the distance between this curve and the position of the minimum along the azimuthal and the radial directions, respectively. Even though this method is only exact for a Gaussian distribution, the results presented in \cref{section_krypton_results} show that these values are a good estimate of the position uncertainty for the smaller S2 pulses.

The uncertainties in the reconstructed positions decrease with increasing number of top array PMTs involved in the reconstruction. However, that improvement is only relevant up to 20 PMTs (including all 59 working channels instead of 20 PMTs improves the position uncertainty by only 3\%). For this reason, we considered in the position reconstruction procedure only the 20 PMTs closest to the initial position estimate (obtained by the corrected centroid algorithm \cite{Solovov2011_PositionReconstruction}). This is done in order to minimize the interference from noise sources (such as the emission of single electrons or afterpulsing in a PMT), which typically lead to events located near the wall of the detector to be reconstructed towards the center. It also improves the speed of the reconstruction by 10\%.

The position reconstruction fails or is affected by a large systematic error when the reconstructed signal is not from a single scatter S2 or the S2 is affected by some source of noise such as  after-pulsing in a PMT or the presence of an unstable PMT etc. Those events have, on average, a larger minimum value of ${q}_{\mathrm{min}}$, and thus a quality cut based on the value of ${q}_{\mathrm{min}}$ can be used to remove those events. In the WS2014--2016, we considered only events with ${q}_{\mathrm{min}}<40+\mathcal{A}_{\mathrm{tot}}/42$. The efficiency of this cut calculated using calibration data (namely ${}^{3}$H) was above 95\% \cite{LUX2016_SSR} for $\mathcal{A}_{\mathrm{tot}}$ between 200 and 4,000~phd.

\subsection{Vertex positions}

The coordinates of the interaction vertex in the $(x, y)$ plane, ($x_{\rm ver}$, $y_{\rm ver}$), coincide with the S2 $(x, y)$ coordinates when a uniform field exists between the cathode and anode of the TPC. In LUX detector, the electric field exhibited some non-uniformity during both  WS2013 and WS2014--16, being more severe and time-dependent in the latter. This is described further in the section 4 and in more detail in \cite{hertel2015, Fields2017}. To obtain the vertex coordinates from the S2 $(x, y)$ position, we created a map $\mathcal{M}$ that converts the reconstructed S2 position into the real vertex positions $x_{\rm ver}$ and $y_{\rm ver}$ 
\begin{equation}
(x_{\rm ver}, y_{\rm ver}) = \mathcal{M}\left(x, y, {\rm drift\,time} \right).
\label{eq_vertex_positions}
\end{equation}
The ${}^{{\rm 83m}}$Kr calibration events, assumed uniformly distributed in the volume of the detector, were used to obtain the map $\mathcal{M}$ and verify its stability in time. This map remained constant along WS2013 but varied along WS2014-2016. A thorough discussion of the creation of those maps is presented in ref. \cite{hertel2015}. It is worth to note that in WS2014-16 the background modeling and the WIMP search data analysis were carried out with the data kept in S2 coordinates, while the true positions of simulated data were mapped into the S2 space using field models developed for this purpose \cite{LUX2016_SSR}. In the following analyses, S2 reconstructed positions are used and not the interaction vertex positions unless stated otherwise.

\section{Results and discussion\label{section_krypton_results}}

The position of each interaction is an essential quantity in the LUX analysis, since both the radius and the azimuthal angle are two of the observables ($r$,  drift time, S1, S2, and $\phi$ in WS2014--16) of the profile likelihood analysis used for the determination of the exclusion limits \cite{LUX2016_SSR}. It is thus crucial to assess the quality of the position reconstruction and determine the uncertainties of the reconstructed positions. As mentioned before, due to the detector design and size, it is not practical to obtain experimental data with known positions of interaction allowing the direct comparison of  the reconstructed positions with the original positions. However, calibration data from dispersed sources have a known distribution of events in the detector. The most useful calibrations for the position reconstruction assessment are the $^{\rm 83m}$Kr (as previously mentioned, used for the construction of the LRFs), the ${}^{3}$H ($\upbeta^-$ with Q~=~18.6~keV), and the 2.45\,MeV D-D neutron calibrations.

In the analyses that follows, WS2013 data are used unless stated otherwise. However, both runs use the same set of LRFs and the same position reconstruction method with no observable changes in the light collection. For the WS2013 results, we observed that $\left<q_{\mathrm{min}}\right>$ was stable within 1\% with no clear degradation observed in the position reconstruction. For the 2014/2016 LUX results \cite{LUX2016_SSR}, a faulty PMT affected the reconstruction  leading to the exclusion of that PMT from the analysis. However, it was sufficient to recover the reconstruction quality without modifying the LRFs. Therefore, the main conclusions presented here are still valid for the WS2014--16 data. The calibration data used here was processed in the same manner as in the analysis of recently published LUX science results \cite{LUX2015_ReanalysisPRL,LUX2016_SSR}.

\subsection{Krypton-83m data\label{section_krypton_calibrations}}

\begin{figure*}
\begin{center}
\includegraphics[width=0.49\textwidth]{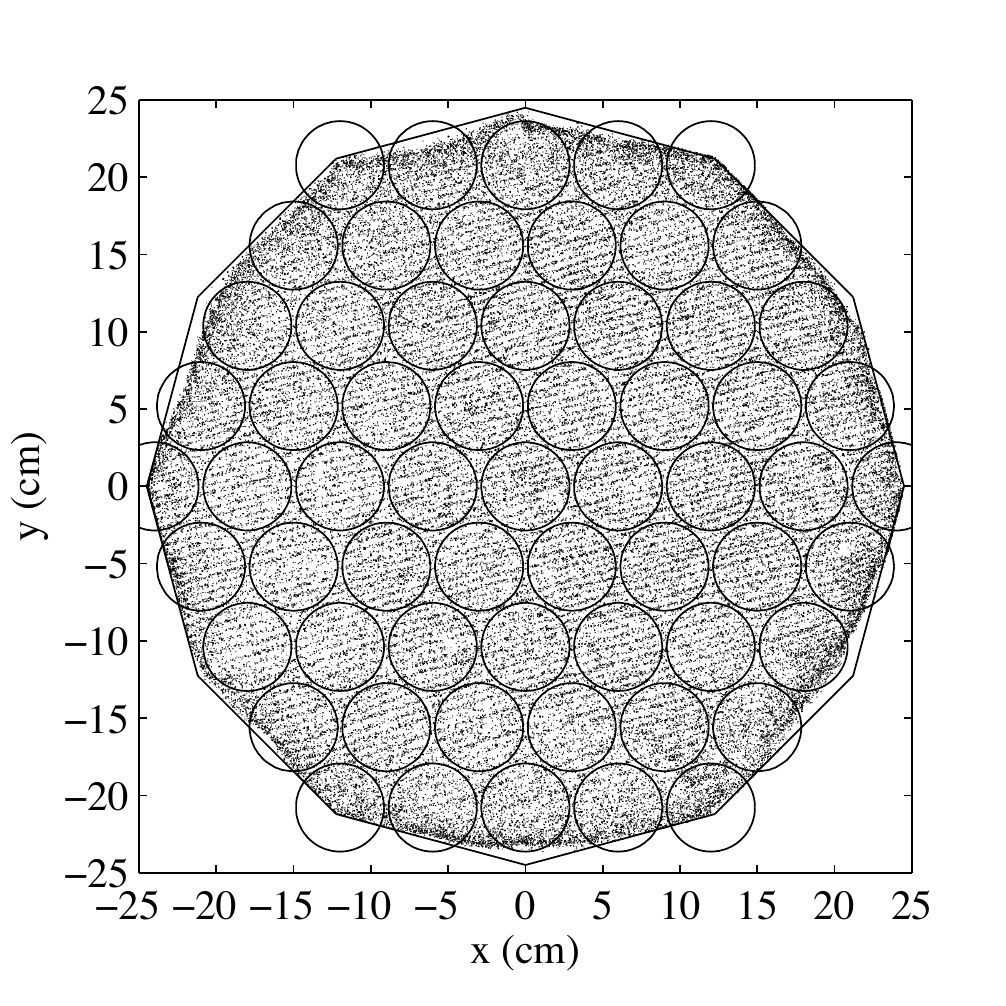}
\includegraphics[width=0.49\textwidth]{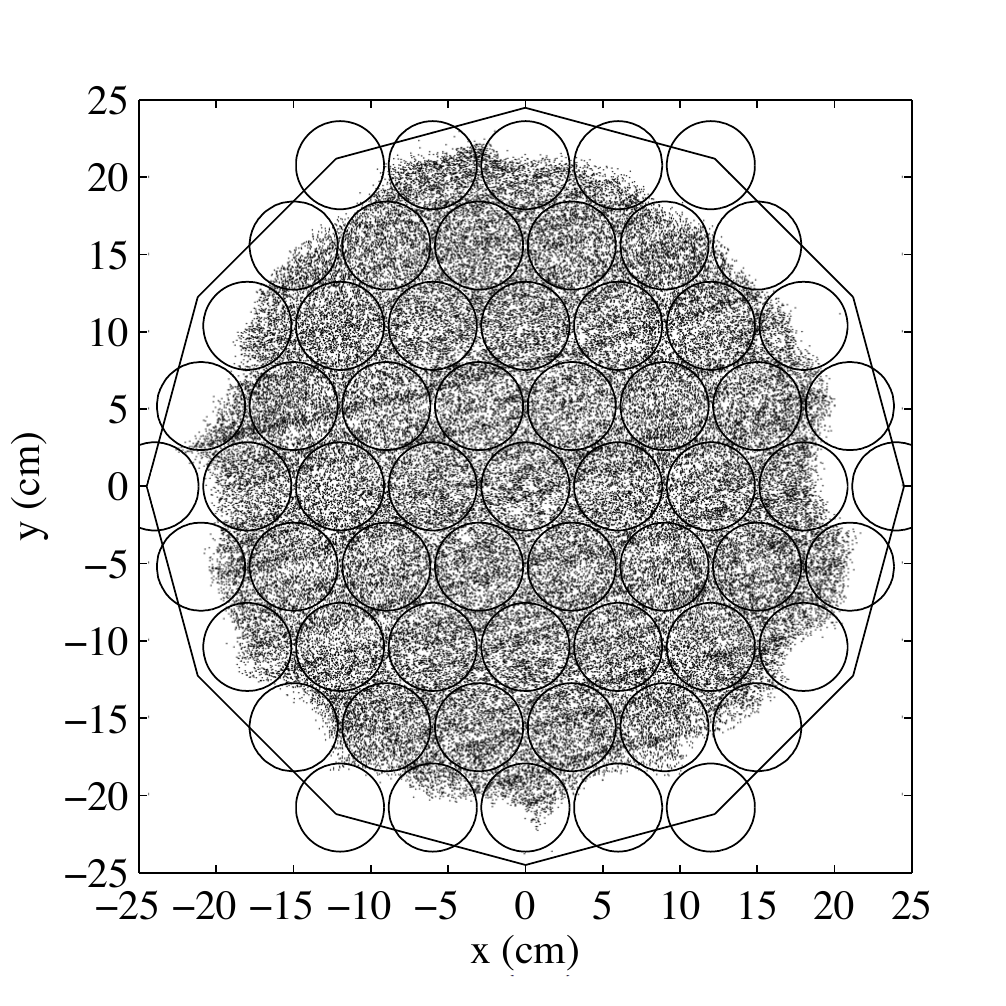}	
\caption{$(x, y)$ scatter plot of the S2 reconstructed positions of $^{\rm 83m}$Kr events for drift time between 4 and 10~$\upmu$s (left panel, WS2014--16) and between 290 and 320~$\upmu$s (right panel, WS2013). The PMT and TPC inner walls are represented by black lines. In both figures, especially at the left, a striped pattern with the pitch of the gate grid wires can be clearly seen.}
\label{Fig14_DensidadeXYCorteScatter}
\end{center}
\end{figure*}

The reconstructed $(x, y)$ distribution of ${}^{{\rm 83m}}$Kr decays in the detector is shown in \cref{Fig14_DensidadeXYCorteScatter} for  events occurring at the top (drift time between 4--10~$\upmu$s) and bottom of the detector (drift time between 290--320~$\upmu$s). From \cref{Fig14_DensidadeXYCorteScatter}, it can be seen that  the reconstructed coordinates do not extend to the edge of the top PMT array. This effect is stronger for longer drift times. Detailed simulations of the electric field in the LUX chamber (described in ref.\ \cite{Fields2017}) explain this observation by the existence of a radial component of the drift field, pushing electrons towards the center of the detector as they drift upwards and hence shifting radially inwards the position of S2 relative to that of the interaction.  In the WS2013, the main origin of the radial field component is the electrical transparency of the cathode grid and  the lateral field rings of the TPC, which allow some field leakage. A nearly identical effect is observed in the XENON100 detector \cite{XENON100_thesis}. This radial field also depends on the azimuthal angle, an effect which is also visible in \cref{Fig14_DensidadeXYCorteScatter}. This may be caused by a combination of azimuthally-varying field leakage across the grid and possibly a non-uniform distribution of accumulated charge on insulating surfaces of the detector. In the WS2014--16 run, the radial field was much stronger  compared with that in the WS2013 run (see ref.\ \cite{Fields2017} for details), but with no direct impact on the LRFs.

\begin{figure}
 \begin{center}
   \includegraphics[width=8.6cm]{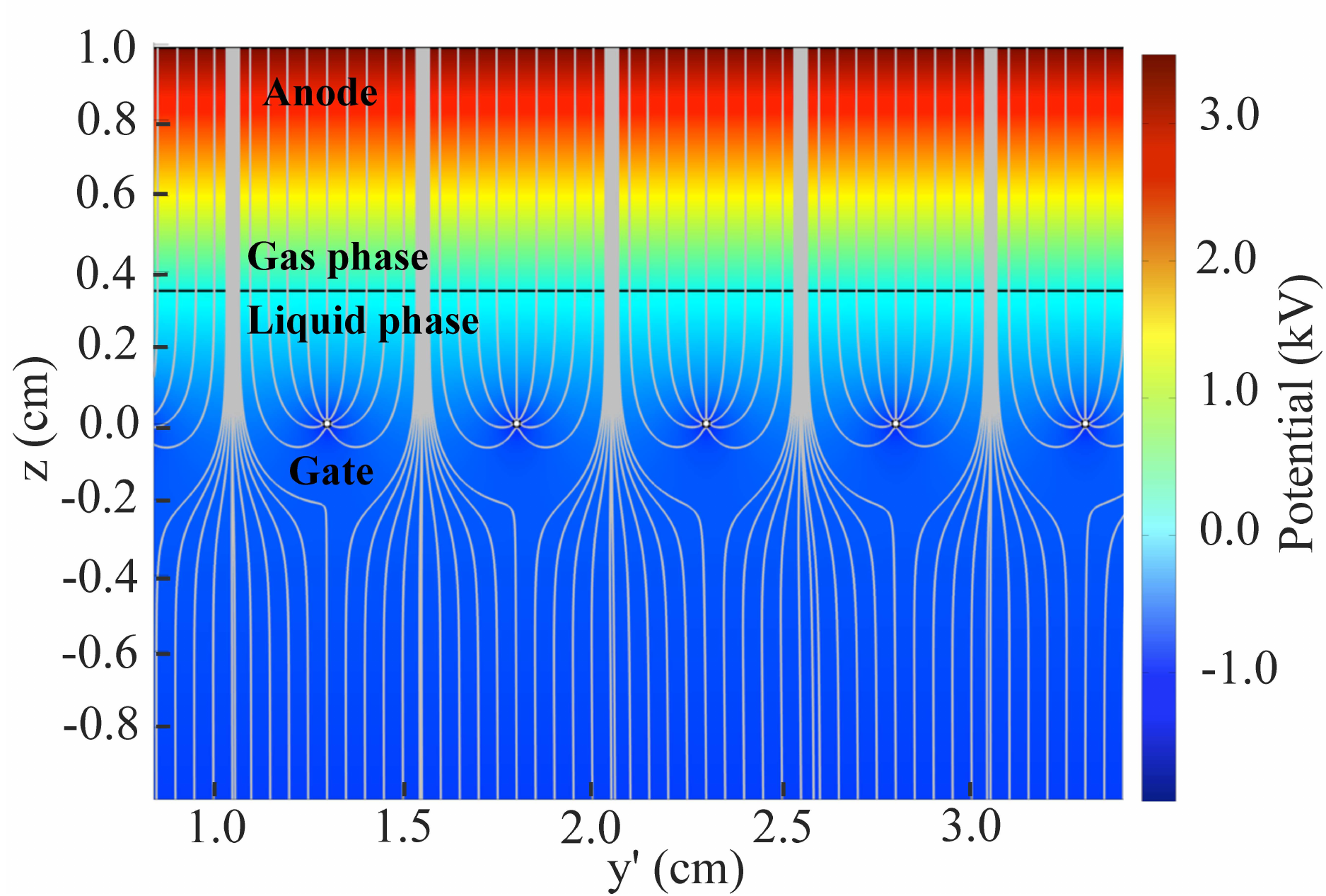}	
\caption{Electron trajectories (in white) and electric potential (color-map) near gate grid wires obtained by a COMSOL Multiphysics simulation \cite{Fields2017}. The gate grid is shown at $z$~=~0~cm by the white dots and the surface of the liquid is represented by the black line. The original $(x, y)$ position of the events were rotated by $15^{\circ}$ in such a way that the direction of the new $x$-axis ($x'$) is parallel to the gate wires, and the direction of the new $y$-axis ($y'$) is perpendicular to the gate wires. Note that the line density is not intended to illustrate the field strength.}
  \label{Fig17_FieldAroundTheGate}
 \end{center}
\end{figure}

In \cref{Fig14_DensidadeXYCorteScatter}, one can observe a striped pattern in the event density, parallel to the wires forming the gate grid (both the gate grid and the observed pattern are at an angle of 15${}^{\circ}$ to the $x$ axis). The  grid wires are 0.1 mm in diameter and 5~mm apart. The grid plane is $\sim$4~mm below the surface of the liquid and 1~cm below the anode grid. This grid separates the drift field (defined by the cathode and gate grids, the former being  48~cm below the gate grid) and the extraction/electroluminescence fields (defined by the gate and anode grids, the latter being 4.8~cm below the top PMTs) in the detector, allowing these two fields to be set independently. Given the large difference between the drift field (180$\pm$20~V/cm in the WS2013) and the extraction field (2.84$\pm$0.16~kV/cm in liquid in WS2013), the drift field lines are compressed as they pass through the gate plane; any electrons leaving the drift volume appear only in narrow strips between each pair of gate wires. This effect is shown in \cref{Fig17_FieldAroundTheGate}, with the path of the electrons along the field lines represented by the white lines and the color-map representing the electric potential. 

\begin{figure}
\begin{center}
\includegraphics[width=8.6cm]{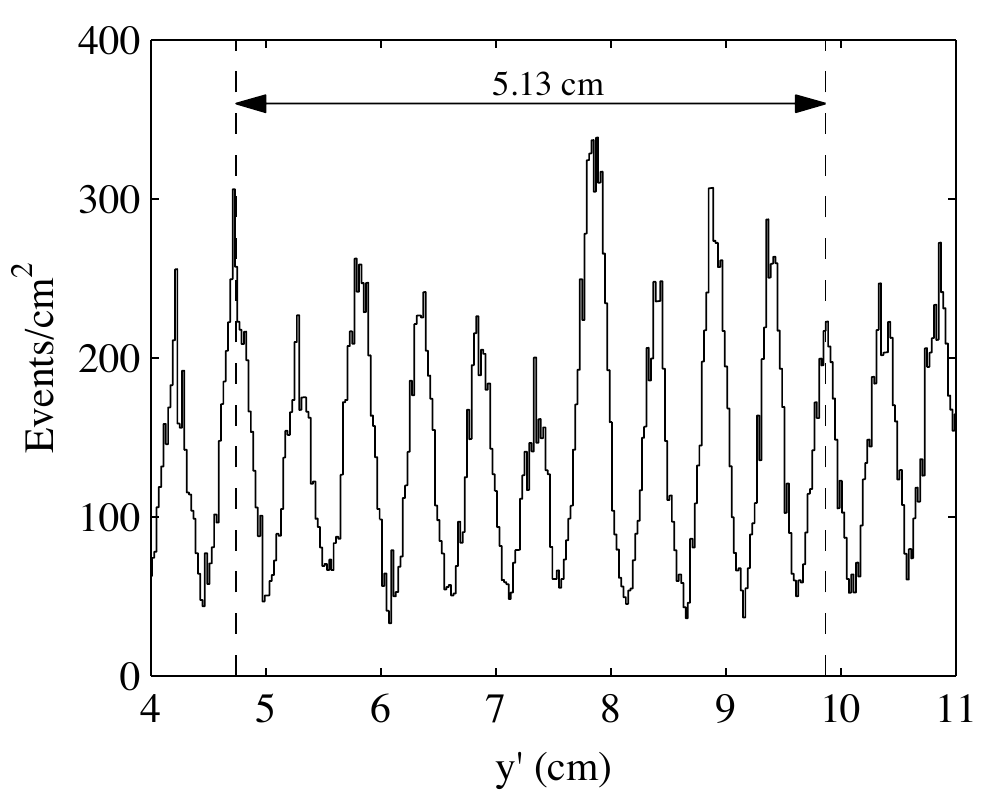}	
\caption{Histogram of the density of ${}^{\mathrm{83m}}$Kr events just below the gate grid represented on an axis perpendicular to the orientation of the gate wires.  Only ${}^{\mathrm{83m}}$Kr events at a depth less than 1 cm below the gate are considered.}
\label{Fig18_DensityGateWires_BelowGate}
\end{center}
\end{figure}

The visualization of this focusing effect of the field lines between the grid wires can be used  to assess the quality of the position reconstruction. The histogram on \cref{Fig18_DensityGateWires_BelowGate} represents the density of the  events occurring right below the gate plane (0 to 1\,cm below the gate) and along the direction perpendicular to the gate wires. Each peak was fitted with a Gaussian distribution to obtain the central position and adjacent strips width. The results show that for the selected region the average distance between two wires is 5.13$\pm$0.07~mm, matching the grid pitch (5~mm) within 2$\sigma$. Additionally, the resolution of the wires determined from the Gaussian width is on average  $\sigma$~=~0.965$\pm$0.028~mm  for the selected region, similar to the average uncertainty from the position reconstruction method ($\sigma$~=~0.86~mm, see \cref{section_uncertainties}). While the strips in the reconstructed position distribution are mostly  uniform right below the gate grid (\cref{Fig14_DensidadeXYCorteScatter} left), large variations in the number of events in each strip are observed in the remaining volume (\cref{Fig14_DensidadeXYCorteScatter} right). These variations are mostly likely to be caused by small variations in the pitch of the wires producing fluctuations of the bulk field which are not observed for the S2 right below the gate wires.  

\begin{figure*}
 \begin{center}
\includegraphics[width=\textwidth]{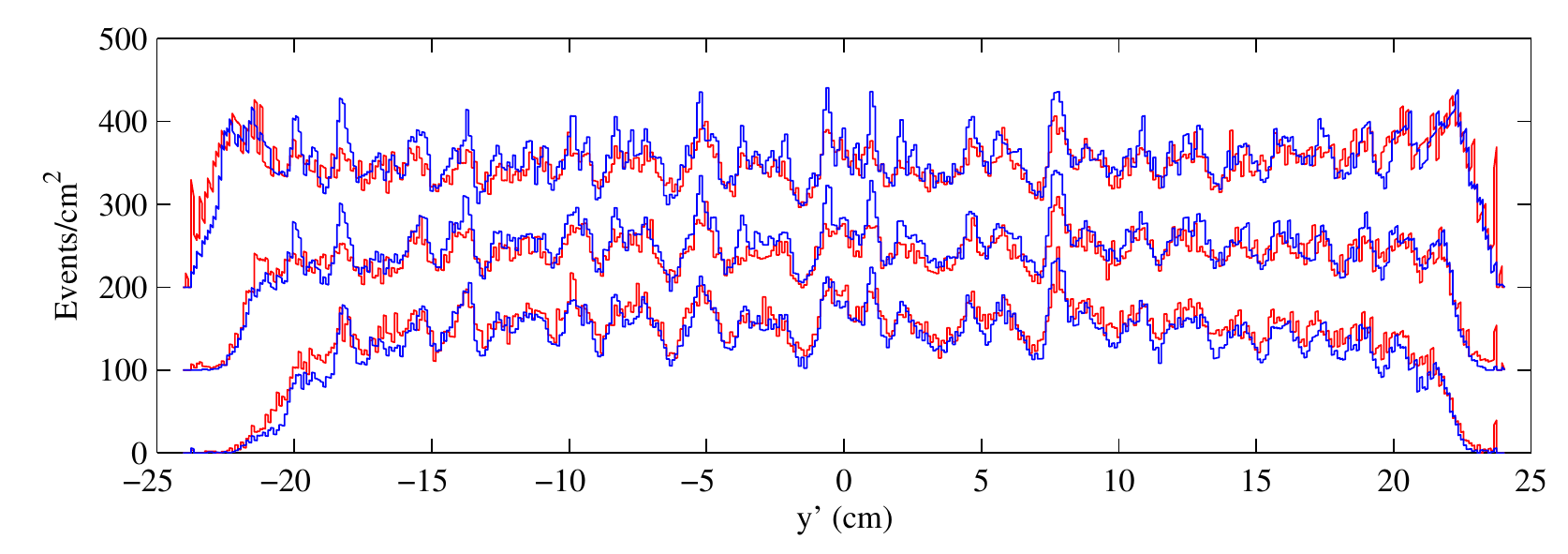}	
\caption{Histogram of the S2 event density for ${}^{3}$H $\upbeta$-decay events (red curves) and ${}^{\mathrm{83m}}$Kr events (blue curves) for different sections of the detector. The two top curves correspond to events from the top third of the chamber (drift time between 5 and 110~$\upmu$s), the two middle curves for events from the middle third of the chamber (drift time between 110 and 215~ $\upmu$s), and the two bottom curves to the bottom third of the chamber (drift time between 215 and 320~$\upmu$s). For clarity, we added a value of +200 to the curves from the top and +100 for the curves from the middle.}
\label{Fig18_DensityGateWires_DriftFieldCut}
\end{center}
\end{figure*}
  
\subsection{Tritium data}

The detector was also calibrated using a tritiated methane source \cite{TritiumPaper2015, AttilaDobi2015}, where the $Q$-value of the ${}^{3}$H $\upbeta^-$ decay is 18.59~keV. We collected more than 300,000 tritium events in a relatively high rate (10~Bq) calibration \cite{TritiumPaper2015}, which is sufficient to compare with ${}^{{\rm 83m}}$Kr calibrations. As in the case of ${}^{\rm 83m}$Kr, the ${}^{3}$H events are expected to be uniformly distributed in the liquid, but the size of the S2 is smaller, up to 6,000~phd. The range of S2 for  ${}^{3}$H is comparable to what is expected from a WIMP event in the search region. For this reason, most of the PMT pulses are described using the photon counting technique instead of pulse area integration as is the case with ${}^{\rm 83m}$Kr. Therefore, ${}^{3}$H is an excellent source to assess the quality of the position reconstruction for smaller S2 pulses and to confirm the consistency of the mixed model described in \cref{section_the_minimization_method}.
 
Similarly to ${}^{{\rm 83m}}$Kr, the distribution of reconstructed tritium events exhibits the same striped pattern in the $(x, y)$ density of events and the presence of the same radial field. \Cref{Fig18_DensityGateWires_DriftFieldCut} shows the event density profiles along the direction perpendicular to the wires for these two sources and for different drift times. As shown, the event density profiles are aligned as expected, showing  no significant systematic error affecting the position reconstruction as function of the drift time and between the two sources. 

\begin{figure}
 \begin{center}
   \includegraphics[width=8.6cm]{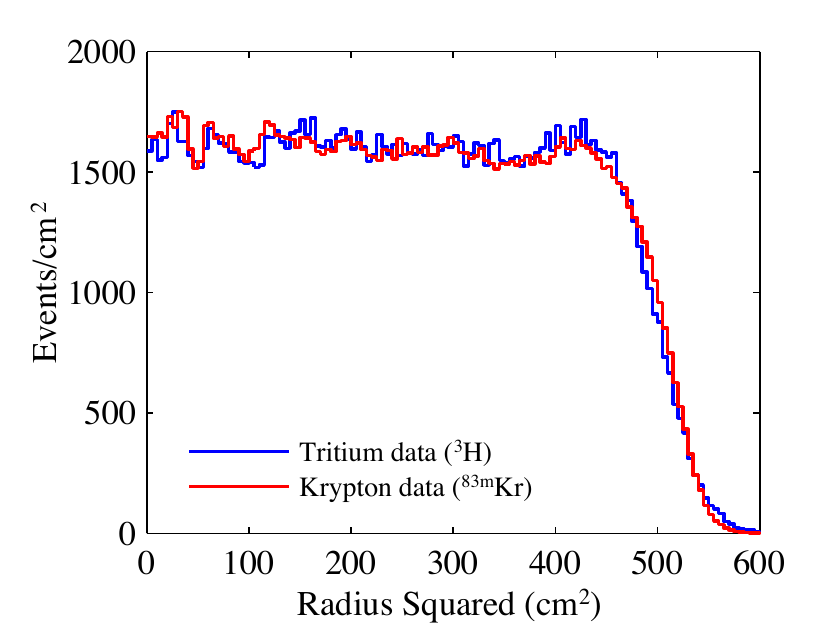}
\caption{Density of events as function of the squared radius for  ${}^{{\rm 83m}}$Kr (red line) and ${}^{3}{\rm H}$ (blue line) data. The event density was normalized at small radii.}
  \label{Fig20_DensityOfEventsKrAndCH3T}
 \end{center}
\end{figure}

Another test performed using tritium data was the comparison between the radial density of ${}^{3}$H and ${}^{{\rm 83m}}$Kr events (\cref{Fig20_DensityOfEventsKrAndCH3T}). The two histograms match well except for a small systematic difference for the events with $r^2$~>~400~cm${}^{2}$. This disagreement is caused by the larger position uncertainties of the tritium events due to their smaller size \cite{LUX2015_ReanalysisPRD}.

\subsection{Neutron data}

\begin{figure}
\begin{center}
\includegraphics[width=8.6cm]{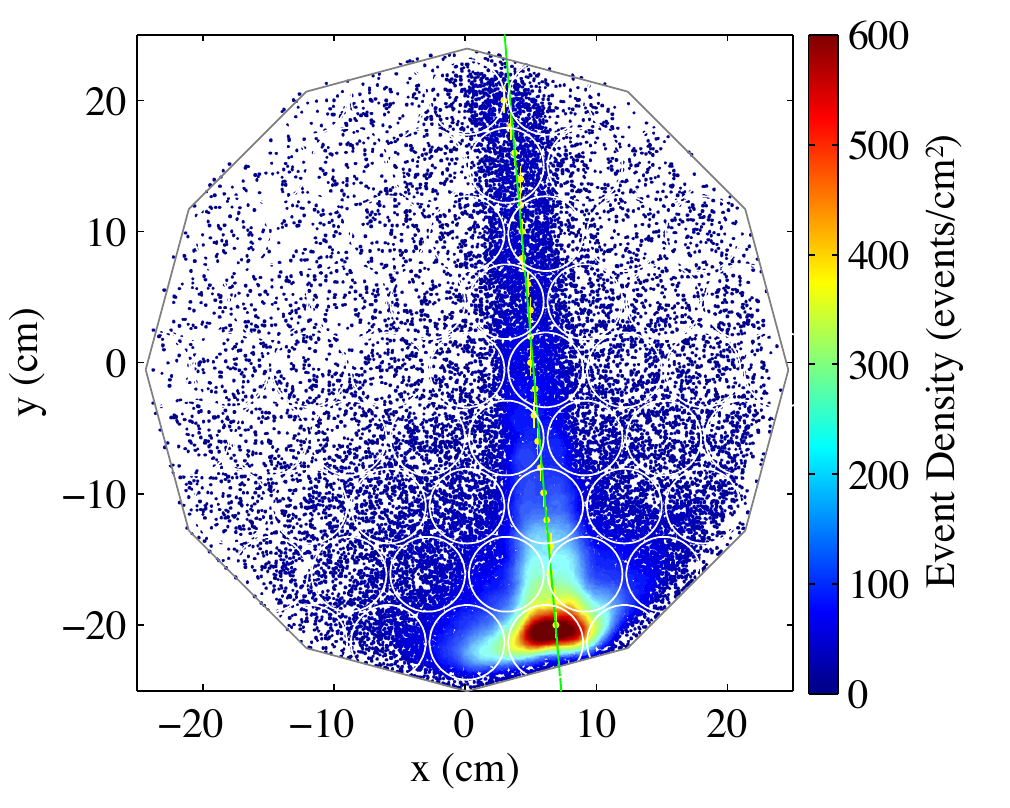}	
\caption{Scatter plot of the reconstructed S2 positions for the neutron data. The colors represent the local $(x,y)$ event density. Yellow points represent the center of the neutron beam measured in 2 cm slices along the neutron beam direction. The x error bars are not visible because they are smaller than the dots, and the yellow y bars correspond to the size of each slice. The green line is the linear fit to the data.}
\label{Fig22_DDCalibrations}
\end{center}
\end{figure}

The response of the detector to nuclear recoils was calibrated  using a D-D neutron generator that emits monoenergetic neutrons with an energy of 2.45~MeV \cite{LUX2016_DDCalibrations, Verbus2017, VerbusPhDThesis2016}. The D-D neutrons are collimated by a 4.9~cm internal diameter air-filled tube which extends between the walls of the shielding water tank and the detector cryostat. Since the position of the first neutron scatter is located in the direction of the beam, the data from these calibrations were used to check for any systematic error affecting the reconstruction.

\Cref{Fig22_DDCalibrations} shows the reconstructed positions of the first neutron scatter.  No correction from the radial field effect was applied to data as, at the height of the neutron beam, this correction is small (only 4\%).

To check the straightness of the reconstructed beam, we divided the beam along the $y$-axis in 2~cm thick slices, and then for each slice we fitted the histogram of the $x$ position to find the center of the beam. The fit is composed of a semi-circle function describing the elastic neutron events inside the beam and a skew-normal distribution describing the background from multiple scatter events and other non-nuclear recoil events. The results are represented in ﬂ\cref{Fig22_DDCalibrations} by the black dots with the error bars corresponding to the uncertainty obtained in the fit plus a systematic contribution from the size and position of the bins. These data are well described by a linear fit represented by the gray line. The root-mean square of the difference between the center of the beam set by the center of the data slices and the fit is only 1.4~mm with the maximum absolute deviation being 4~mm.

\subsection{Position uncertainties and resolution\label{section_uncertainties}}

The LUX position reconstruction code calculates two statistical position uncertainties ($\sigma_r$, $\sigma_\phi$) in the reconstructed positions (\cref{section_Likelihood_Ratio_Analysis}). These uncertainties are used in: i) the neutron calibrations, to estimate the uncertainty in the scattering angle of double scatters \cite{LUX2016_DDCalibrations}; and ii) calculating the wall-event contribution to the background model, to estimate the contamination of wall events in the fiducial volume \cite{LUX2016_SSR}.

\begin{figure}
 \begin{center}
   \includegraphics[width=8.5cm]{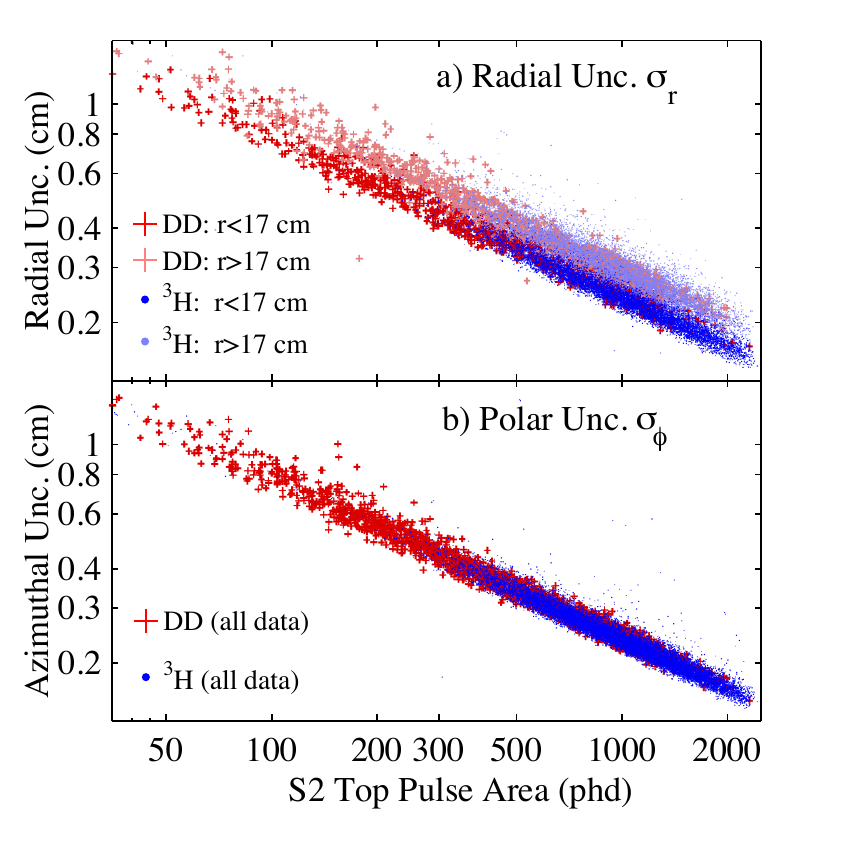}%
\caption{ Average position uncertainties $\sigma_r$ (panel a) and $\sigma_\phi$ (panel b) as a function of the pulse area in the top array $\mathcal{A}_{\rm top}$. The data are from tritium (blue dots) and D-D neutron (red crosses) calibrations.  In LUX, the S2 size is estimated using $\mathcal{A}_{T}$ which includes information from both arrays.}
  \label{Fig21_ErrorsCH3andDD}
 \end{center}
\end{figure}

\Cref{Fig21_ErrorsCH3andDD} shows both uncertainties as a function of the  S2 area in the top array, $\mathcal{A}_{\rm top}$ ($\left<\mathcal{A}_{\rm top}/\mathcal{A}_{T}\right>$~=~0.566$\pm$0.019 for $r$~<~20~cm). This figure includes calibration data from both tritium (blue dots) and D-D nuclear recoil events (red crosses), for a reconstructed radius smaller than 17~cm (dark red/blue markers) and larger than 17~cm (light red/blue markers). As shown, the uncertainties depend mostly on $\sigma^2\propto1/\mathcal{A}_{\rm top}$ reflecting the Poisson distribution of the photon statistics.  

\begin{figure}
 \begin{center}
\begin{overpic}[width=0.5\textwidth,tics=10]{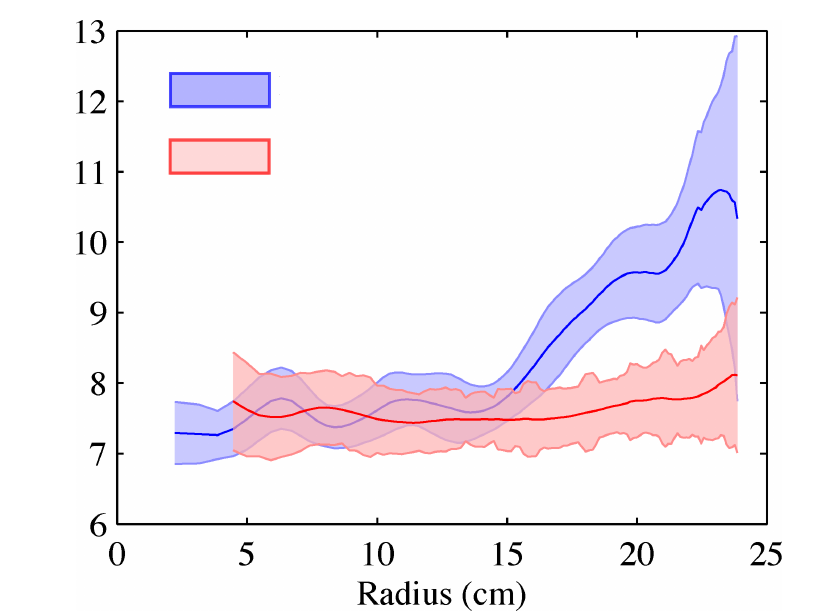}
 \put(-1.0,11){\rotatebox{90}{\footnotesize $\Upsilon_{r,\phi}$ $\left(\sigma_{r,\phi}\sqrt{\mathcal{A}_{\rm top}}\right)$, $\left(\mathrm{cm}\cdot\sqrt{\mathrm{phd}}\right)$}}
 \put(34,61.5){\rotatebox{0}{\footnotesize $\Upsilon_r$, ${}^{3}$H data, 1-$\sigma$ range}}
 \put(34,53){\rotatebox{0}{\footnotesize $\Upsilon_\phi$, ${}^{3}$H data, 1-$\sigma$ range}}

\end{overpic}

\caption{Average position uncertainties $\Upsilon_r$ ($\sigma_r\sqrt{\mathcal{A}_{\rm top}}$, blue curves) and $\Upsilon_{\phi}$ ($\sigma_\phi\sqrt{\mathcal{A}_{\rm top}}$, red curves) as a function of the reconstructed radius $r$. The colored regions correspond to the 1-$\sigma$ range on the distribution of the position uncertainties.}
  \label{FigExtra12_ChangeOfKWithR}

 \end{center}
\end{figure}

We can normalize the uncertainties to the number of photons, defining $\Upsilon_r$ and $\Upsilon_{\phi}$ as
\begin{equation}
	\Upsilon_{(r,\phi)} = \sigma_{(r,\phi)} {\sqrt{\mathcal{A}_{\rm top}}}. \label{uncerteq}
\end{equation}
The average values of $\Upsilon_r$ and $\Upsilon_\phi$ as a function of the radius $r$ are represented on \cref{FigExtra12_ChangeOfKWithR}. As the figure confirms, $\Upsilon_r$ has a significant radial dependence, being about 40\% larger for peripheral events than for central ones.
Conversely, $\Upsilon_\phi$ is mostly constant. Additionally, the values of $\Upsilon_r$ and $\Upsilon_\phi$ depend on the distance of the event to the center of the nearest PMT. We fobserved that both uncertainties are on average 8\% larger for S2 pulses generated right below the center of a PMT than for those near the border of a PMT. We can explain this by looking to the first derivative of the radial component, $\left|\eta'\right|$, (see figure \ref{Fig06_RadialComponent_Data}) which is maximized for $\rho$ between 3 and 5~cm. The number of PMTs that fall into that region is larger when the light is emitted near the border of a PMT.

\begin{figure}
 \begin{center}
   \includegraphics{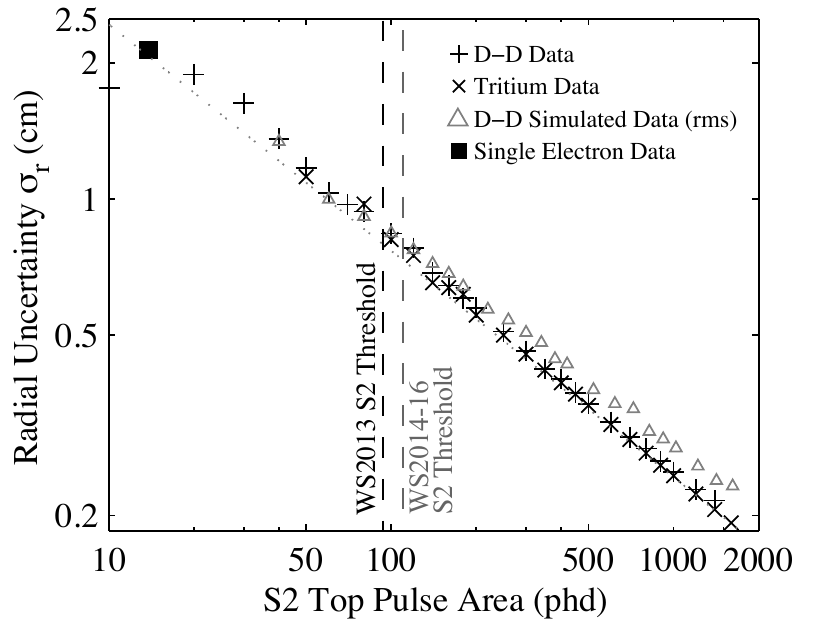}	
\caption{Radial statistical uncertainty, $\sigma_r$, as a function of the pulse area of the top PMT array. The data are from the neutron calibrations, tritium calibrations, single electrons from ${}^{\textrm{83m}}$Kr, and simulations of D-D neutron calibrations, covering a wide range of pulse areas. For the simulations, the uncertainty corresponds to the root mean square of the difference between the reconstructed radius and the true radial position of the event. For both data and simulations, only the events with a reconstructed radius smaller than 20~cm and a drift time between 40 and 300~$\upmu$s are shown. The uncertainty for single electron events is $\sigma_r$~=~2.13~cm (average top pulse area of 13.8~phd). The black and gray dashed vertical lines corresponds to the S2 threshold used in the WS2013 analysis (165~phd) and in the WS2014--16 analysis (200~phd) respectively. The dotted line corresponds to a fit on the form $K\sqrt{\mathcal{A}_{\mathrm{top}}}$ simultaneously to both D-D and tritium data.}
  \label{Fig22_UncertainsComparison}
 \end{center}
\end{figure}

The comparison of the experimental uncertainties with those obtained from simulated data is shown in \cref{Fig22_UncertainsComparison}. It shows the average value of the uncertainties obtained with  tritium, neutron data and single electrons emitted after the ${}^{\mathrm{83m}}$Kr S2 signal, as well as with simulation data of neutron calibrations. The data are for a reconstructed radius smaller than 20~cm. The uncertainties follow a square-root dependence as predicted by photon statistics. For the simulated data, the uncertainty corresponds to the root mean square of the difference between the reconstructed radius and the true radial position of the event. There is only a small difference ($\sim$0.5~mm) between the uncertainties obtained from experimental data and from simulations. For $r < 20$~cm, the radial uncertainty is 0.93~cm for the WS2013 S2 threshold ($\mathcal{A}_T$ = 165~phd, $\mathcal{A}_{\mathrm{top}}$~=~93.4~phd, <~1~keV$_{\mathrm{nr}}$) and 0.82~cm for the WS2014--16 S2 threshold ($\mathcal{A}_T$ = 200~phd, $\mathcal{A}_{\mathrm{top}}$~=~113.2~phd). These values were obtained with the 2013 D-D data. The uncertainty decreases to 0.17~cm for $\mathcal{A}_T$ = 4,000~phd  ($\mathcal{A}_{\mathrm{top}}$~=~2,264~phd, $\sim$10~keV$_{\mathrm{ee}}$) obtained with the 2013 ${}^{3}$H data.

An S2 due to a single electron that escapes from the liquid into the gas is the smallest electroluminescence signal that can be observed  ($\mathcal{A}_{T}\simeq$22~phd, \cref{Fig02_Mercury_PCVsPA}); thus, the study of the single electron events is paramount to assess the reconstruction of very small signals and to determine the lower limit of the position resolution. In a detector such as LUX, the single electrons are typically caused by \cite{EDWARDS200854, Santos2011}:
\begin{enumerate}
\item{delayed single emission from a previous S2 event: electrons can accumulate under the surface barrier at the liquid/gas interface and escape into the gas later};
\item{photoionization of impurities in the liquid by the S2 or S1 photons}.  
\end{enumerate}
The delayed electrons (1) are expected to have the same reconstructed $(x, y)$ position of the parent S2 signal, so they can be used to directly measure the position resolution for single electrons. On the contrary, single electrons from photoionization (2) are created anywhere along the path of the S2 (or S1) light in the liquid being almost uncorrelated with the position of the parent S2.

In this study, we selected single electrons emitted after the S2 pulse of ${}^{{\rm 83m}}$Kr events. The uncertainty in the $(x, y)$ position of the ${}^{{\rm 83m}}$Kr S2 pulse is negligible compared with that of the single electron since it is much larger (between 4,000 and 20,000 phd). This  selection of single electrons includes both single electrons from delayed emission and photoionization. To reduce the background from the latter, only single electrons observed within a time window, $\Delta\tau$, less than $\Delta\tau$~<~20~$\upmu$s after the end of the ${}^{{\rm 83m}}$Kr S2 signal were accepted.

\begin{figure}
 \begin{center}
  \includegraphics[width=8.5cm]{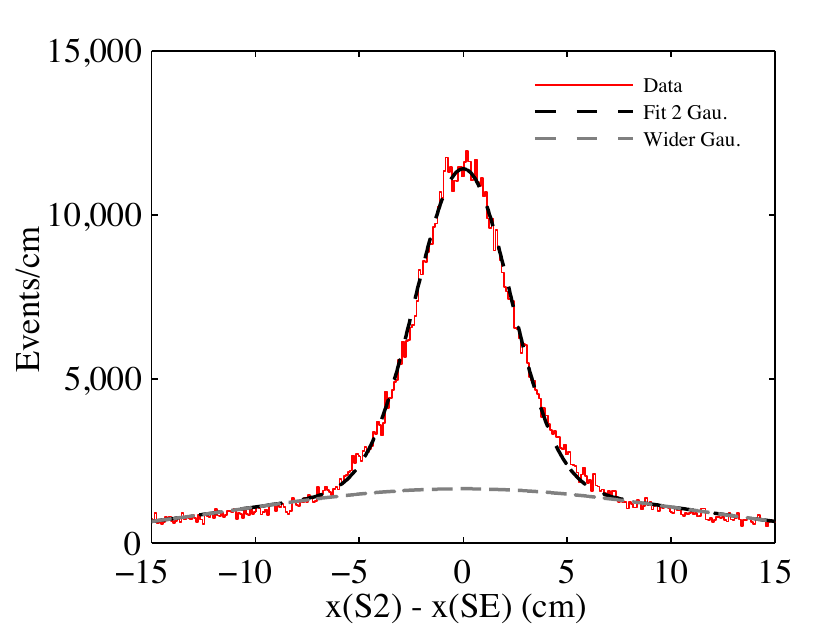}
  \caption{Difference between the reconstructed $x$ position of a S2 ${}^{\mathrm{83m}}$Kr signal, $x$(S2), and the reconstructed $x$ position of the single electron observed after the ${}^{\mathrm{83m}}$Kr signal, $x$(SE). Only single electrons signals with a total pulse area from 10 to 35 phd, $r$~<~20~cm, and $\Delta\tau$~<~20~$\upmu$s are analyzed. The dashed black curve corresponds to the fit with the sum of two Gaussians centered at zero, and the dashed gray curve corresponds to the contribution from the broader Gaussian distribution.}
  \label{Fig29_ErroPosicao_PulseAreaCentro}
 \end{center}
\end{figure}

The distribution of the difference between the reconstructed $x$ position of the ${}^{\mathrm{83m}}$Kr S2 signal and the reconstructed $x$ position of the associated single electron, $\Delta x$, is shown in \cref{Fig29_ErroPosicao_PulseAreaCentro}. This histogram is fitted with the sum of two Gaussians with different widths, both centered at zero. 
The narrower distribution corresponds to the single electrons from the delayed emission while the wider Gaussian describes the contribution from photoionization single electrons. The resolution of the single electrons given by the standard deviation of the narrower Gaussian distribution was determined to be 2.24$\pm$0.04~cm, comparable to the average uncertainty in the x position (2.13~cm) obtained for these single electrons in the position reconstruction. In contrast, the standard deviation of the wider Gaussian is about 11~cm. The results for the $\Delta y$ match exactly with the results for $\Delta x$.

The contamination of the single electrons from photoionization can be estimated by measuring the relative area of the wider Gaussian distribution. For our selection of $\Delta\tau$~<~20~$\upmu$s, the photoionization accounts for 46.8\% of all the single electrons. For larger values of $\Delta\tau$, the number of single electrons from delayed emission as function of $\Delta\tau$ follows an exponential decay law with a decay time constant of $\sim$40\,$\upmu$s.  

We checked the existence of any possible drift of single electrons from delayed emission along the liquid surface of the detector as this would deteriorate the position resolution. The existence of such drift was tested by measuring the average value of $\Delta x$ and $\Delta y$ as function of $\Delta\tau$. For this, the data were sliced in 20~$\upmu$s sections in $\Delta\tau$ up to a maximum value of 200~$\upmu$s and for each of those slices the $\Delta x$ and $\Delta y$ histograms were fitted  using a double Gaussian  with the mean of both Gaussian distributions not fixed. We observed that for $\Delta\tau$~<~200~$\upmu$s the absolute value of the fitted mean is always smaller than 2~mm for both $\Delta x$ and $\Delta y$, and no significant increase in the standard deviation of the distribution was observed. This is a good indication that there is no significant movement of the charge along the surface of the liquid.

We performed this same analysis using only the pulse areas to reconstruct the position of single electrons instead of the mixed method adopted through this paper (both described in \cref{subsection_pulse_areas}). The resolution obtained in the double Gaussian fit was 2.47$\pm$0.04~cm which corresponds to a degradation on the resolution of single electrons of 10.27$\pm$0.25\%.  This shows that the mixed method employed here improves the reconstruction of low energy events.

The uncertainties of the true vertex positions, ($x_{\rm ver}$, $y_{\rm ver}$), are affected by additional contributions. As before, the uncertainties are described using cylindrical coordinates ($\sigma_r^{\rm ver}$, $\sigma_\phi^{\rm ver}$). The uncertainty along the radial direction $\sigma_r^{\rm ver}$ is obtained in the following way:
\begin{equation}
\left(\sigma_r^{\rm ver}\right)^2 = \left(\frac{r_{\rm ver}}{r} \sigma_r \right)^2 + \sigma_{\rm gate}^2 + \sigma_{\rm sys}^2 \left(x_{\rm ver}, y_{\rm ver} \right),
\label{equncertaintyvertex}
\end{equation}
 where $r_{\rm ver}$ corresponds to the vertex radial position.
For the azimuthal uncertainty, $\sigma_\phi^{\rm ver}$, the first term is replaced by $\sigma_\phi^2$. The first term in \cref{equncertaintyvertex} corresponds to the error propagation of the  uncertainties in the S2 position reconstruction,  $\sigma_{\left(r,\phi\right)}$. $r_{\rm ver}/r$ measures the effect of the stretching along the radial direction. For the WS2013 data, $\left<r_{\rm ver}/r\right>$ is ~1.0 in the top, $~$1.06 in the middle, and $~$1.15 in the bottom of the chamber. For the WS2014--16 data, $\left<r_{\rm ver}/r\right>$ increases to about 3 in the bottom of the chamber. The second term, $\sigma_{\rm gate}$, corresponds to the contribution of the gate focusing effect to the final uncertainty. We estimated this contribution by taking the standard deviation of the distribution of the distances between events uniformly distributed in $(x, y)$ to the nearest wire. From this, we obtained $\sigma_{\rm gate} = $~1.1~mm, independent of the event depth. The third term, $\sigma_{\rm sys}$, is the systematic uncertainty associated to the process of obtaining the vertex coordinates from the reconstructed S2 position using \cref{eq_vertex_positions}. This uncertainty was estimated directly from the data using two different methods: i)  study of  wall events produced in the radiative decay of both ${}^{\mathrm{210}}$Pb and its daughters and ii) study of the neutron D-D beam linearity. Using the population of wall events, we estimated $\sigma_{\rm sys}$ to be 1~mm for WS2013 and between 2 and 3~mm for the WS2014-16 for events close to the walls. The study of the linearity of the D-D beam in corrected variables limited $\sigma_{\rm sys}$ to a maximum of 4~mm for events in the center of the chamber (see \cite{LUX2015_ReanalysisPRD}).

\section{Summary}

A statistical method was developed to obtain the $(x, y)$ position of an interaction in the LUX detector from the observed S2 signal distribution at the top PMT array. This method employs \emph{in situ} calibrations to obtain the LRF for each photomultiplier. The presence of PTFE reflectors around the sensitive volume increases the complexity of the LRFs. These were written as the sum of two terms: an axial component $\eta\left(\rho\right)$ describing the light that goes directly to the PMTs or is reflected in the liquid surface and a polar component $\varepsilon\left(\rho\right)$ characterizing the light reflected from the PTFE walls. 

We assessed the quality of the position reconstruction using calibration data obtained during LUX detector operations. Although in these data the position of each individual interaction is not known, the resolution can be measured using both the effect of the focusing effect of the electrons going through the gate grid and the single electrons emitted right after a ${}^{{\rm 83m}}$Kr S2 event. This resulted in an observed resolution of 0.0965$\pm$0.0028\,cm for the ${}^{{\rm 83m}}$Kr S2 pulses (with average areas of 22,000~phd) and 2.24$\pm$0.04\,cm for single electrons (with average areas of 22~phd), which agree well with the calculated uncertainties obtained in the position reconstruction algorithm (0.086~cm for ${}^{{\rm 83m}}$Kr S2 pulses and 2.13~cm for single electrons). 

The analysis of systematic errors was done by looking to the uniformity in $(x, y)$ of both ${}^{{\rm 3}}$H and ${}^{{\rm 83m}}$Kr reconstructed events, the linearity of the neutron beam produced in the D-D neutron calibrations, and by measuring the average distance between two wires of the gate grid. All these tests verified the absence of  significant systematic uncertainties affecting the position reconstruction.

For the position reconstruction of very small S2s ($\mathcal{A}_T$~<~4,000~phd), we employed a method where the response of a PMT is described using photon counting or pulse areas according to the pulse size observed in the PMT. This method was  faster when compared to a pure maximum likelihood method based only on pulse areas. Moreover, we observed no significant systematic error on the low S2 pulse data (${}^{3}$H and single electrons) when compared with the large S2 pulse (${}^{{\rm 83m}}$Kr), and the analysis of the position resolution of single electrons obtained with the mixed method revealed an improvement of 10.27$\pm$0.25\% when compared with a pure pulse areas method. 

\acknowledgments

This work was partially supported by the U.S. Department of Energy (DOE) under award numbers DE-AC02-05CH11231, DE-AC05-06OR23100, DE-AC52-07NA27344, DE-FG01-91ER40618, DE-FG02-08ER41549, DE-FG02-11ER41738, DE-FG02-91ER40674, DE-FG02-91ER40688, DE-FG02-95ER40917, DE-NA0000979, DE-SC0006605, DE-SC0010010, and DE-SC0015535; the U.S. National Science Foundation under award numbers PHY-0750671, PHY-0801536, PHY-1003660, PHY-1004661, PHY-1102470, PHY-1312561, PHY-1347449, PHY-1505868, and PHY-1636738; the Research Corporation grant RA0350; the Center for Ultra-low Background Experiments in the Dakotas (CUBED); and the South Dakota School of Mines and Technology (SDSMT). LIP-Coimbra acknowledges funding from Funda\c{c}\~{a}o para a Ci\^{e}ncia e a Tecnologia (FCT) through the project-grant PTDC/FIS-NUC/1525/2014. Imperial College and Brown University thank the UK Royal Society for travel funds under the International Exchange Scheme (IE120804). The UK groups acknowledge institutional support from Imperial College London, University College London and Edinburgh University, and from the Science \& Technology Facilities Council for PhD studentships ST/K502042/1 (AB), ST/K502406/1 (SS) and ST/M503538/1 (KY). The University of Edinburgh is a charitable body, registered in Scotland, with registration number SC005336.

We gratefully acknowledge the logistical and technical support and the access to laboratory infrastructure provided to us by SURF and its personnel at Lead, South Dakota. SURF was developed by the South Dakota Science and Technology Authority, with an important philanthropic donation from T. Denny Sanford, and is operated by Lawrence Berkeley National Laboratory for the Department of Energy, Office of High Energy Physics.

\bibliographystyle{JHEP}
\bibliography{Reconstruction}

\providecommand{\href}[2]{#2}\begingroup\raggedright\begin{thebibliography}{10}

\bibitem{LUX_2013NIM}
{\scshape LUX} collaboration, D.~Akerib et~al., \emph{The large underground
  xenon ({LUX}) experiment},
  \href{http://dx.doi.org/http://dx.doi.org/10.1016/j.nima.2012.11.135}{\emph{Nucl.
  Instr. Meth. Phys. Res. A} {\bfseries 704} (2013) 111},
  [\href{https://arxiv.org/abs/1211.3788}{{\ttfamily 1211.3788}}].

\bibitem{Chen2012}
W.~T. Chen, H.~Carduner, J.~P. Cussonneau, J.~Donnard, S.~Duval, A.~F.
  Mohamad-Hadi et~al., \emph{Measurement of the transverse diffusion
  coefficient of charge in liquid xenon},  in \emph{Proceedings in Diffusion in
  Solids and Liquids VII, Algarve, Portugal}, vol.~326 of \emph{Defect and
  Diffusion Forum}, p.~567, Trans Tech Publications, 2012.
\newblock
  \href{http://dx.doi.org/10.4028/www.scientific.net/DDF.326-328.567}{DOI}.

\bibitem{PhysRevC.95.025502}
{\scshape EXO} collaboration, J.~B. Albert et~al., \emph{Measurement of the
  drift velocity and transverse diffusion of electrons in liquid xenon with the
  {EXO}-200 detector},
  \href{http://dx.doi.org/10.1103/PhysRevC.95.025502}{\emph{Phys. Rev. C}
  {\bfseries 95} (Feb, 2017) 025502},
  [\href{https://arxiv.org/abs/1609.04467}{{\ttfamily 1609.04467}}].

\bibitem{LUX2015_BackgroundPaper}
{\scshape LUX} collaboration, D.~Akerib et~al., \emph{Radiogenic and
  muon-induced backgrounds in the {LUX} dark matter detector},
  \href{http://dx.doi.org/http://dx.doi.org/10.1016/j.astropartphys.2014.07.009}{\emph{Astropart.
  Phys.} {\bfseries 62} (2015) 33},
  [\href{https://arxiv.org/abs/1403.1299}{{\ttfamily 1403.1299}}].

\bibitem{ChangLeePhDThesis}
C.~Lee, \emph{Mitigation of Backgrounds for the {Large Underground Xenon} Dark
  Matter Detector}.
\newblock PhD thesis, Case Western Reserve University, 2015.

\bibitem{Fields2017}
{\scshape LUX} collaboration, D.~S. Akerib et~al., \emph{{{3D} Modeling of
  Electric Fields in the {LUX} Detector}},
  \href{http://dx.doi.org/10.1088/1748-0221/12/11/P11022}{\emph{J. Instrum.}
  {\bfseries 12} (2017) P11022},
  [\href{https://arxiv.org/abs/1709.00095}{{\ttfamily 1709.00095}}].

\bibitem{LUX2016_DDCalibrations}
{\scshape LUX} collaboration, D.~S. Akerib et~al., \emph{{Low-energy (0.7-74
  keV) nuclear recoil calibration of the LUX dark matter experiment using D-D
  neutron scattering kinematics}}, {\emph{ArXiv e-prints} (2016) },
  [\href{https://arxiv.org/abs/1608.05381}{{\ttfamily 1608.05381}}].

\bibitem{Anger1958_ScintillationCamera}
H.~O. Anger, \emph{Scintillation camera},
  \href{http://dx.doi.org/http://dx.doi.org/10.1063/1.1715998}{\emph{Rev. Sci.
  Instrum.} {\bfseries 29} (1958) 27}.

\bibitem{Short_1984}
M.~Short, \emph{{Proceedings of the International Workshop on X- and
  $\gamma$-Ray Imaging Techniques Gamma-camera systems}},
  \href{http://dx.doi.org/http://dx.doi.org/10.1016/0167-5087(84)90192-3}{\emph{Nucl.
  Instr. Meth. Phys. Res.} {\bfseries 221} (1984) 142}.

\bibitem{Morozov2016}
A.~Morozov, V.~Solovov, R.~Martins, F.~Neves, V.~Domingos and V.~Chepel,
  \emph{{ANTS2} package: simulation and experimental data processing for
  {Anger} camera type detectors},
  \href{http://dx.doi.org/10.1088/1748-0221/11/04/P04022}{\emph{J. Instrum.}
  {\bfseries 11} (2016) P04022},
  [\href{https://arxiv.org/abs/1602.07247}{{\ttfamily 1602.07247}}].

\bibitem{Pelssers_2015MT}
B.~Pelssers, \emph{Position reconstruction and data quality in {XENON}},
  Master's thesis, Universiteit Utrecht, 2015.

\bibitem{Gray1976}
R.~M. Gray and A.~Macovski, \emph{Maximum a posteriori estimation of position
  in scintillation cameras},
  \href{http://dx.doi.org/10.1109/TNS.1976.4328354}{\emph{IEEE Trans. Nucl.
  Sci.} {\bfseries 23} (Feb, 1976) 849--852}.

\bibitem{Lindote2007}
A.~Lindote, H.~Ara{\'u}jo, J.~P. da~Cunha, D.~Akimov, V.~Chepel, D.~Davidge
  et~al., \emph{Preliminary results on position reconstruction for
  {ZEPLIN-III}},
  \href{http://dx.doi.org/https://doi.org/10.1016/j.nima.2006.10.254}{\emph{Nucl.
  Instr. Meth. Phys. Res. A} {\bfseries 573} (2007) 200}.

\bibitem{Solovov2011_PositionReconstruction}
V.~N. Solovov, V.~A. Belov, D.~Y. Akimov, H.~M. Ara{\'u}jo, E.~J. Barnes, A.~A.
  Burenkov et~al., \emph{Position reconstruction in a dual phase xenon
  scintillation detector},
  \href{http://dx.doi.org/10.1109/NSSMIC.2011.6154607}{\emph{NSS/MIC, 2011
  IEEE, Valencia, Spain} (2011) 1226},
  [\href{https://arxiv.org/abs/1112.1481}{{\ttfamily 1112.1481}}].

\bibitem{LUX2014_OriginalResults}
{\scshape LUX} collaboration, D.~S. Akerib et~al., \emph{First results from the
  {LUX} dark matter experiment at the {Sanford Underground Research Facility}},
  \href{http://dx.doi.org/10.1103/PhysRevLett.112.091303}{\emph{Phys. Rev.
  Lett.} {\bfseries 112} (2014) 091303},
  [\href{https://arxiv.org/abs/1310.8214}{{\ttfamily 1310.8214}}].

\bibitem{LUX2015_ReanalysisPRL}
{\scshape LUX} collaboration, D.~S. Akerib et~al., \emph{Improved limits on
  scattering of weakly interacting massive particles from reanalysis of 2013
  {LUX} data},
  \href{http://dx.doi.org/10.1103/PhysRevLett.116.161301}{\emph{Phys. Rev.
  Lett.} {\bfseries 116} (2016) 161301},
  [\href{https://arxiv.org/abs/1512.03506}{{\ttfamily 1512.03506}}].

\bibitem{LUX2016SpinDependent}
{\scshape LUX} collaboration, D.~S. Akerib et~al., \emph{Results on the
  spin-dependent scattering of weakly interacting massive particles on nucleons
  from the {Run 3} data of the {LUX} experiment},
  \href{http://dx.doi.org/10.1103/PhysRevLett.116.161302}{\emph{Phys. Rev.
  Lett.} {\bfseries 116} (2016) 161302},
  [\href{https://arxiv.org/abs/1602.03489}{{\ttfamily 1602.03489}}].

\bibitem{LUX2017_Axions}
{\scshape LUX} collaboration, D.~S. Akerib et~al., \emph{First searches for
  axions and axionlike particles with the {LUX} experiment},
  \href{http://dx.doi.org/10.1103/PhysRevLett.118.261301}{\emph{Phys. Rev.
  Lett.} {\bfseries 118} (2017) 261301},
  [\href{https://arxiv.org/abs/1704.02297}{{\ttfamily 1704.02297}}].

\bibitem{LUX2016_SSR}
{\scshape LUX} collaboration, D.~S. Akerib et~al., \emph{Results from a search
  for dark matter in the complete {LUX} exposure},
  \href{http://dx.doi.org/10.1103/PhysRevLett.118.021303}{\emph{Phys. Rev.
  Lett.} {\bfseries 118} (2017) 021303},
  [\href{https://arxiv.org/abs/1608.07648}{{\ttfamily 1608.07648}}].

\bibitem{LUX2017_SD}
{\scshape LUX} collaboration, D.~S. Akerib et~al., \emph{Limits on
  spin-dependent {WIMP}-nucleon cross section obtained from the complete lux
  exposure},
  \href{http://dx.doi.org/10.1103/PhysRevLett.118.251302}{\emph{Phys. Rev.
  Lett.} {\bfseries 118} (2017) 251302},
  [\href{https://arxiv.org/abs/1705.03380}{{\ttfamily 1705.03380}}].

\bibitem{hertel2015}
{\scshape LUX} collaboration, D.~S. Akerib et~al., \emph{{$^{83\textrm{m}}$Kr
  calibration of the 2013 {LUX} dark matter search}}, {\emph{Phys. Rev. D
  (forthcoming)} (2017) }, [\href{https://arxiv.org/abs/1708.02566}{{\ttfamily
  1708.02566}}].

\bibitem{TritiumPaper2015}
{\scshape LUX} collaboration, D.~S. Akerib et~al., \emph{Tritium calibration of
  the {LUX} dark matter experiment},
  \href{http://dx.doi.org/10.1103/PhysRevD.93.072009}{\emph{Phys. Rev. D}
  {\bfseries 93} (2016) 072009},
  [\href{https://arxiv.org/abs/1512.03133}{{\ttfamily 1512.03133}}].

\bibitem{Darkside2015_FirstResults}
{\scshape DarkSide} collaboration, \emph{First results from the {DarkSide}-50
  dark matter experiment at {Laboratori} {Nazionali} del {Gran} {Sasso}},
  \href{http://dx.doi.org/http://dx.doi.org/10.1016/j.physletb.2015.03.012}{\emph{Phys.
  Lett.} {\bfseries 743} (2015) 456},
  [\href{https://arxiv.org/abs/1410.0653}{{\ttfamily 1410.0653}}].

\bibitem{PandaX_2016}
{\scshape PandaX-II} collaboration, A.~Tan et~al., \emph{Dark matter results
  from first 98.7 days of data from the {PandaX-II} experiment},
  \href{http://dx.doi.org/10.1103/PhysRevLett.117.121303}{\emph{Phys. Rev.
  Lett.} {\bfseries 117} (2016) 121303},
  [\href{https://arxiv.org/abs/1607.07400}{{\ttfamily 1607.07400}}].

\bibitem{Morozov2015_PosRec}
A.~Morozov, V.~Solovov, F.~Alves, V.~Domingos, R.~Martins, F.~Neves et~al.,
  \emph{Iterative reconstruction of detector response of an {Anger} gamma
  camera}, {\emph{Phys. Med. Biol.} {\bfseries 60} (2015) 4169}.

\bibitem{Jortner1965}
J.~Jortner, L.~Meyer, S.~A. Rice and E.~G. Wilson, \emph{Localized excitations
  in condensed {Ne}, {Ar}, {Kr}, and {Xe}},
  \href{http://dx.doi.org/http://dx.doi.org/10.1063/1.1695927}{\emph{J. Chem.
  Phys.} {\bfseries 42} (1965) 4250}.

\bibitem{FUJII2015293}
K.~Fujii, Y.~Endo, Y.~Torigoe, S.~Nakamura, T.~Haruyama et~al.,
  \emph{High-accuracy measurement of the emission spectrum of liquid xenon in
  the vacuum ultraviolet region},
  \href{http://dx.doi.org/http://dx.doi.org/10.1016/j.nima.2015.05.065}{\emph{Nucl.
  Instr. Meth. Phys. Res. A} {\bfseries 795} (2015) 293}.

\bibitem{LUX2013_PMTs}
{\scshape LUX} collaboration, D.~Akerib et~al., \emph{An ultra-low background
  {PMT} for liquid xenon detectors},
  \href{http://dx.doi.org/http://dx.doi.org/10.1016/j.nima.2012.11.020}{\emph{Nucl.
  Instr. Meth. Phys. Res. A} {\bfseries 703} (2013) 1},
  [\href{https://arxiv.org/abs/1205.2272}{{\ttfamily 1205.2272}}].

\bibitem{FranciscoNeves2017_Reflectancia}
F.~Neves, A.~Lindote, A.~Morozov, V.~Solovov, C.~Silva, P.~Bras et~al.,
  \emph{Measurement of the absolute reflectance of polytetrafluoroethylene
  {(PTFE)} immersed in liquid xenon}, {\emph{J. Instrum.} {\bfseries 12} (2017)
  P01017}.

\bibitem{LUX2015_ReanalysisPRD}
{\scshape LUX} collaboration, D.~S. Akerib et~al., \emph{{Calibration, event
  reconstruction, data analysis and limits calculation for the LUX dark matter
  experiment}}, {\emph{arXiv e-print} (2017) },
  [\href{https://arxiv.org/abs/1712.05696}{{\ttfamily 1712.05696}}].

\bibitem{PhysRevC.80.045809}
L.~W. Kastens, S.~B. Cahn, A.~Manzur and D.~N. McKinsey, \emph{Calibration of a
  liquid xenon detector with $^{\mathrm{83m}}\mathrm{Kr}$},
  \href{http://dx.doi.org/10.1103/PhysRevC.80.045809}{\emph{Phys. Rev. C}
  {\bfseries 80} (2009) 045809}.

\bibitem{LUXDAQ2012}
{\scshape LUX} collaboration, D.~S. Akerib et~al., \emph{Data acquisition and
  readout system for the {LUX} dark matter experiment},
  \href{http://dx.doi.org/http://dx.doi.org/10.1016/j.nima.2011.11.063}{\emph{Nucl.
  Instr. Meth. Phys. Res. A} {\bfseries 668} (2012) 1},
  [\href{https://arxiv.org/abs/1108.1836}{{\ttfamily 1108.1836}}].

\bibitem{nicodemus}
F.~Nicodemus, J.~Richmond, J.~Hsia, I.~Ginsberg and T.~Limperis,
  \emph{Geometrical Considerations and Nomenclature for Reflectance}.
\newblock National Bureau of Standards, 1977.

\bibitem{Claudio2007}
C.~Silva, J.~P. da~Cunha, V.~Chepel, A.~Pereira, V.~Solovov, P.~Mendes et~al.,
  \emph{Measuring the angular profile of the reflection of xenon scintillation
  light}, \href{http://dx.doi.org/DOI: 10.1016/j.nima.2007.05.166}{\emph{Nucl.
  Instr. Meth. Phys. Res. A} {\bfseries 580} (2007) 322}.

\bibitem{silva:064902}
C.~Silva, J.~P. da~Cunha, A.~Pereira, V.~Chepel, M.~I. Lopes, V.~Solovov
  et~al., \emph{Reflectance of polytetrafluoroethylene for xenon scintillation
  light}, \href{http://dx.doi.org/10.1063/1.3318681}{\emph{J. Appl. Phys.}
  {\bfseries 107} (2010) 064902}.

\bibitem{LUX2012_LUXSim}
{\scshape LUX} collaboration, D.~Akerib et~al., \emph{{LUXSim:} a
  component-centric approach to low-background simulations},
  \href{http://dx.doi.org/http://dx.doi.org/10.1016/j.nima.2012.02.010}{\emph{Nucl.
  Instr. Meth. Phys. Res. A} {\bfseries 675} (2012) 63},
  [\href{https://arxiv.org/abs/1111.2074}{{\ttfamily 1111.2074}}].

\bibitem{Cowan1998}
G.~Cowan, \emph{Statistical Data Analysis}.
\newblock Oxford science publications. Clarendon Press, 1998.

\bibitem{Macovski_1976}
R.~M. Gray and A.~Macovski, \emph{Maximum a posteriori estimation of position
  in scintillation cameras},
  \href{http://dx.doi.org/10.1109/TNS.1976.4328354}{\emph{IEEE Trans. Nucl.
  Sci.} {\bfseries 23} (1976) 849}.

\bibitem{logLikelihood}
T.~J. Cleophas and A.~H. Zwinderman, \emph{Log likelihood ratio tests},  in
  \emph{Statistical Analysis of Clinical Data on a Pocket Calculator}, p.~37.
\newblock Springer Netherlands, 2011.
\newblock \href{http://dx.doi.org/10.1007/978-94-007-1211-9_13}{DOI}.

\bibitem{Wilks1938_LikelihoodRatio}
S.~S. Wilks, \emph{The large-sample distribution of the likelihood ratio for
  testing composite hypotheses},
  \href{http://dx.doi.org/10.1214/aoms/1177732360}{\emph{Ann. Math. Statist.}
  {\bfseries 9} (1938) 60}.

\bibitem{5402300}
S.~Vinogradov, T.~Vinogradova, V.~Shubin, D.~Shushakov and K.~Sitarsky,
  \emph{Probability distribution and noise factor of solid state
  photomultiplier signals with cross-talk and afterpulsing},  in \emph{IEEE
  Nucl. Sci. Symp. Conf. Rec. 2009}, p.~1496, Oct, 2009.
\newblock \href{http://dx.doi.org/10.1109/NSSMIC.2009.5402300}{DOI}.

\bibitem{LUXCF_thesis}
C.~H. Faham, \emph{Prototype, Surface Commissioning and Photomultiplier Tube
  Characterization for the Large Underground Xenon ({LUX}) Direct Dark Matter
  Search Experiment}.
\newblock Ph.D. dissertation, Brown University, 2014.

\bibitem{PDG2014}
K.~Olive and P.~D. Group, \emph{Review of particle physics - chapter 32
  statistics}, {\emph{Chin. Phys. C} {\bfseries 38} (2014) 090001}.

\bibitem{Faham2015}
C.~Faham, V.~Gehman, A.~Currie, A.~Dobi, P.~Sorensen and R.~Gaitskell,
  \emph{Measurements of wavelength-dependent double photoelectron emission from
  single photons in {VUV}-sensitive photomultiplier tubes}, {\emph{J. Instrum.}
  {\bfseries 10} (2015) P09010},
  [\href{https://arxiv.org/abs/1506.08748}{{\ttfamily 1506.08748}}].

\bibitem{Sorensen2011}
P.~Sorensen, \emph{Anisotropic diffusion of electrons in liquid xenon with
  application to improving the sensitivity of direct dark matter searches},
  \href{http://dx.doi.org/http://dx.doi.org/10.1016/j.nima.2011.01.089}{\emph{Nucl.
  Instr. Meth. Phys. Res. A} {\bfseries 635} (2011) 41},
  [\href{https://arxiv.org/abs/1102.2865}{{\ttfamily 1102.2865}}].

\bibitem{XENON100_thesis}
Y.~Mei, \emph{Direct Dark Matter Search with the XENON100 Experiment}.
\newblock PhD thesis, Rice University, 2011.

\bibitem{AttilaDobi2015}
A.~Dobi, \emph{Measurement of the Electron Recoil Band of the LUX Dark Matter
  Detector With a Tritium Calibration Source}.
\newblock PhD thesis, University of Maryland, 2014.

\bibitem{Verbus2017}
J.~Verbus, C.~Rhyne, D.~Malling, M.~Genecov, S.~Ghosh, A.~Moskowitz et~al.,
  \emph{Proposed low-energy absolute calibration of nuclear recoils in a
  dual-phase noble element {TPC} using {D-D} neutron scattering kinematics},
  \href{http://dx.doi.org/http://dx.doi.org/10.1016/j.nima.2017.01.053}{\emph{Nucl.
  Instr. Meth. Phys. Res. A} {\bfseries 851} (2017) 68},
  [\href{https://arxiv.org/abs/1608.05309}{{\ttfamily 1608.05309}}].

\bibitem{VerbusPhDThesis2016}
J.~Verbus, \emph{An Absolute Calibration of Sub-1~keV Nuclear Recoils in Liquid
  Xenon Using D-D Neutron Scattering Kinematics in the {LUX} Detector}.
\newblock PhD thesis, Brown University, 2016.

\bibitem{EDWARDS200854}
B.~Edwards, H.~Ara{\'u}jo, V.~Chepel, D.~Cline, T.~Durkin, J.~Gao et~al.,
  \emph{Measurement of single electron emission in two-phase xenon},
  \href{http://dx.doi.org/http://dx.doi.org/10.1016/j.astropartphys.2008.06.006}{\emph{Astropart.
  Phys.} {\bfseries 30} (2008) 54}.

\bibitem{Santos2011}
E.~Santos, B.~Edwards, V.~Chepel, H.~M. Ara{\'u}jo, D.~Y. Akimov, E.~J. Barnes
  et~al., \emph{Single electron emission in two-phase xenon with application to
  the detection of coherent neutrino-nucleus scattering},
  \href{http://dx.doi.org/10.1007/JHEP12(2011)115}{\emph{J. High Energy Phys.}
  {\bfseries 2011} (2011) 115},
  [\href{https://arxiv.org/abs/1110.3056}{{\ttfamily 1110.3056}}].

\end{thebibliography}\endgroup

\end{document}